\documentclass[twocolumn,preprintnumbers,amsmath,amssymb]{revtex4}
\usepackage{amsmath,amssymb,graphics,epsfig,subfigure}
\usepackage{color}
\usepackage{blindtext}
\usepackage{booktabs}

\usepackage{etoolbox} 
\usepackage{lipsum} 
\usepackage[capitalize]{cleveref}

\makeatletter
\appto{\appendix}{%
  \@ifstar{\def\theequation@prefix{A.}}%
          {}%
}
\makeatother

\usepackage{graphicx}

\begin{document}

\thispagestyle{empty}

\begin{center}

\title{Contact and metric structures in black hole chemistry}

\date{\today}
\author{Aritra Ghosh\footnote{E-mail: ag34@iitbbs.ac.in} and Chandrasekhar Bhamidipati\footnote{E-mail: chandrasekhar@iitbbs.ac.in}}

\affiliation{School of Basic Sciences, Indian Institute of Technology Bhubaneswar,\\   Jatni, Khurda, Odisha, 752050, India}

\begin{abstract}
We review recent studies of contact and thermodynamic geometry for black holes in AdS spacetimes in the extended thermodynamics framework. The cosmological constant gives rise to the notion of pressure $P = -\Lambda / 8 \pi$ and, subsequently a conjugate volume $V$, thereby leading to a close analogy with hydrostatic thermodynamic systems. To begin with, we 
review the contact geometry approach to thermodynamics in general and then consider thermodynamic metrics constructed as the Hessians of various thermodynamic potentials. We then study their correspondence to statistical ensembles for systems with two-dimensional spaces of equilibrium states. From the zeroes and divergences of the curvature scalar obtained from the metric, we carefully analyze the issue of ensemble non-equivalence and show certain complimentary behaviors in the description of a thermodynamic system. Following a thorough analysis of the familiar van der Waals system, we turn our attention to black holes in extended phase space. Considering the example of charged AdS black holes, we discuss the generic features of their thermodynamic geometry in detail. The relationship of the thermodynamic curvature(s) with critical points as well as microscopic interactions in black holes is also briefly explored. We finally set up the thermodynamic geometry for finite temperature gauge theories dual to black holes in AdS via holographic correspondence and comment on recent progress.

\end{abstract}

\maketitle
\end{center}


\section{Introduction}
It is well appreciated that black holes are associated with entropy and temperature, being given by the relations \cite{Bardeen:1973gs,Bekenstein:1973ur,Bekenstein:1974ax,Hawking:1974sw,Hawking:1976de} (in units \(\hbar = k_B = c = G = 1\)):
\begin{equation}
S = \frac{\mathcal{A}}{4}, \hspace{7mm} T = \frac{\kappa}{2 \pi} 
\end{equation} where \(\mathcal{A}\) is the area of the event horizon whereas, \(\kappa\) is the surface gravity at the horizon. The mass of the black hole \(M\) can be interpreted as the fundamental energy function whose variations satisfy: \(\delta M = T \delta S + \mu_i \delta C^i\) where \(C^i\) are typically the conserved charges such as electric charge and angular momentum, while \(\mu_i\) are relevant chemical potentials. Asymptotically flat space black holes are typically not thermodynamically stable at non-zero temperatures. However, black holes in AdS, i.e. with a negative cosmological constant (\(\Lambda < 0 \)) can reach thermodynamic stability via the Hawking-Page transition \cite{Hawking:1982dh}. The study of asymptotically AdS black holes is further motivated from the AdS/CFT correspondence \cite{Maldacena:1997re,Witten:1998qj,Witten:1998zw,Chamblin,Chamblin:1999hg}, wherein the Hawking-Page transition corresponds to a confinement-deconfinement transition on the boundary \cite{Witten:1998zw}. In the recent years, it has been shown that the cosmological constant \(\Lambda\) can be treated as a thermodynamic pressure, leading to novel pressure-volume variables in black hole mechanics \cite{Kastor:2009wy,Dolan:2010ha,Cvetic:2010jb}. Thus, the first law of black hole thermodynamics gets modified to:  \(\delta M = T \delta S + V \delta P + \mu_i \delta C^i\), where \(P = -\Lambda/8 \pi\) \cite{Kastor:2009wy}, while \(V\) is known as the thermodynamic volume \cite{Cvetic:2010jb} appearing as conjugate to \(P\). Such a viewpoint, has lead to a flurry of research over the past decade, and a close analogy between the thermodynamics of black holes in AdS and that of ordinary hydrostatic systems has been discovered \cite{Kubiznak:2012wp,Gunasekaran:2012dq,GBcriticality,lower} (see also \cite{Johnson,DolanBose,Karch}). Remarkably, phase transitions of black holes have been shown to be closely analogous to the liquid-gas phase transition, exhibited by the van der Waals model, with an exact matching of the critical exponents \cite{Kubiznak:2012wp}. \\

The introduction of geometrical ideas into thermodynamics have led to several interesting physical insights (see for example, the reviews \cite{Rup1,Aman}). Of particular interest is the concept of a length between different thermodynamic states \cite{Rup2,Weinhold,Crooks}. A good starting point leading to the notion of thermodynamic length is Einstein's fluctuation theory which can be motivated as follows. It is a well understood fact that the entropy of a thermodynamic system is a measure of the number of ways the system can arrange itself microscopically. One can then invert the Boltzmann's formula for entropy (we shall set \(k_B = 1\) throughout the paper): \(S = \ln \Omega\) where \(\Omega\) is the thermodynamic probability or equivalently the number of accessible microstates to obtain
\begin{equation}\label{Omega}
  \Omega = e^S.
\end{equation} We may then expand the entropy \(S\) about its equilibrium value \(S_0\), i.e. about the point at which all its first derivatives vanish so that we have up to the second order:
\begin{equation}
 S \approxeq S_0 + \frac{\partial^2 S_0}{\partial x^i \partial x^j} dx^i \otimes dx^j
\end{equation} where \(\{x^i\}\) are suitable thermodynamic variables specified by external baths or boundary conditions defining the ensemble. With this, one can re-write eqn (\ref{Omega}) as
\begin{equation}\label{OmegaS}
  \Omega \sim e^{-dl_R^2}
\end{equation} with,
\begin{equation}\label{dlR}
  dl_R^2 = - \frac{\partial^2 S_0}{\partial x^i \partial x^j} dx^i \otimes dx^j.
\end{equation}
Clearly, \(dl_R^2\) can be interpreted as a length on the space of thermodynamic equilibrium states between points \(x^i\) and \(x^i + dx^i\). With this, eqn (\ref{OmegaS}) can be interpreted as follows: the shorter the length is between two thermodynamic states, the more probable is a fluctuation between them! It therefore follows from elementary fluctuation theory that the notion of a length on the spaces of thermodynamic equilibrium states is very well motivated physically. The metric given in eqn (\ref{dlR}) for suitable thermodynamic variables \(\{x^i\}\) is called the Ruppeiner metric \cite{Rup1,Rup2}. Since the second law of thermodynamics implies that the entropy of a system is a concave function, the Ruppeiner metric is positive definite. \\

The curvature scalar associated with the Ruppeiner metric, known as the Ruppeiner curvature or simply the thermodynamic curvature possesses an intriguing behavior. The empirical understanding obtained from studying several thermodynamic systems is as follows. It typically diverges at critical points and possibly also at points where the thermodynamic system exhibits strong microscopic correlations. This has been verified for several systems including the van der Waals fluid \cite{Wei1,Wei2} and model magnetic systems \cite{Ising,Dolan:1997cf,Dolan:2002wm} (see also \cite{Kenna,exponent,information}). In fact, it has been argued \cite{Rup1,Rup3} that close to the critical point, the Ruppeiner curvature scales with the correlation volume, i.e. \(R \sim \xi^d\) where \(\xi\) is the correlation length and \(d\) is the number of spatial dimensions. Another interesting aspect of the Ruppeiner curvature is that its sign seems to have a connection with the nature of dominant interactions between the microscopic degrees of freedom in a given thermodynamic system \cite{Rup3}. In the sign convention that we adopt in this paper, the curvature scalar is negative (\(R < 0\)) for the attractive van der Waals gas \cite{Wei1,Wei2} or an ideal gas of bosons \cite{quantum}. In the latter, the attractive interactions are of quantum mechanical origin. Similarly, for the ideal gas of fermions, one has \(R > 0\) which may be taken to signal the existence of quantum mechanical repulsive interactions whose origin can be traced back into the exclusion principle \cite{quantum}. This feature has indeed been verified for several systems where independent microscopic calculations can be performed (see \cite{Rup3} and references therein). Therefore, the Ruppeiner curvature seems to be a powerful diagnostic tool whose behavior may reveal early insights into the microscopic physics of systems such as black holes where a satisfactory microscopic theory is not yet available \cite{Wei1,Wei2,Cai,Aman:2003ug,Shen:2005nu,Sarkarbtz,Mirza:2007ev,Quevedo:2007mj,Hendi:2015rja,Mansoori:2016jer,Banerjee:2010da,Sahay:2010tx,Liu:2010sz,AR,Wei:2015iwa,Bhattacharya:2017hfj,Miao:2017fqg,Xu,weiGB,meanfield1,GBfixedphi,wei4DGB,mansoori4DGB,btz,btz2,meanfield2,Dehyadegari:2020ebz,HosseiniMansoori:2020jrx,unicrit,NaveenaKumara:2020biu,Wei:2020kra,Yerra:2020tzg} (see also \cite{chemical1,chemical2,chemical3,chemical4,chemical5}).\\

In recent times, there has been an ongoing debate about the applicability of geometric methods for understanding the physics of thermodynamic systems \cite{con1,con2,con3,con4}, including black holes. Thermodynamic or information geometry has been shown to be a powerful diagnostic tool, with the divergences of the associated curvature scalar capturing the critical points in various thermodynamic systems. The connection between thermodynamic curvature and the specific heat capacities has also been explored, particularly because the divergences of these quantities typically signal the onset of instabilities and phase transitions in a system. Despite these advantages, there have been a few longstanding unresolved issues concerning the lack of diffeomorphism invariance~\cite{Quevedo:2007mj,geometrotherm} and non-equivalence of thermodynamic curvatures constructed out of different thermodynamic potentials~\cite{bravettiensemble}, among others. It has been noted in several papers in the past that thermodynamic Hessian metrics are not Legendre invariant (see for example \cite{Quevedo:2007mj,bravettiensemble,Nettel}) and we can understand it by considering thermodynamic curvatures constructed from two different thermodynamic potentials. Without loss of generality, for example, one can construct a certain thermodynamic curvature $R_U$ by taking the internal energy $U$ as the fundamental potential with the divergences of $R_U$ capturing the critical point of the system. Performing a partial Legendre transform and using instead enthalpy $H$ as the potential, leads to a different thermodynamic curvature (call it, $R_H$), which may not capture the critical point exactly (as we show later). Thus, the features exhibited by thermodynamic curvatures computed using different potentials may be different. This has been termed as ensemble non-equivalence in thermodynamic geometry \cite{bravettiensemble}. While a large body of work is devoted to probing the connection of the divergences of thermodynamic curvatures to phase transitions, there has been a relatively less focus on the study of their zeroes, until recently~\cite{Wei1,Wei2,meanfield1,btz,meanfield2}. \\

One of the major goals of this article is to pedagogically introduce thermodynamic geometry with black holes in mind, and to perform a critical analysis of the thermodynamic curvatures obtained in different statistical ensembles related by (partial) Legendre transforms. It is demonstrated that if $R_U$ captures the divergences of a certain specific heat, $R_H$ (obtained by a Legendre transform) contains information about the zeroes of that specific heat and vice versa. This suggests a complimentary nature of \(R_U\) and \(R_H\) (see also \cite{bravettiensemble}). Different parametrizations of thermodynamic Hessian metrics in a given ensemble are elaborately discussed. We follow the general route discussed by Mrugala \cite{Mmetricpaper} (discussed in section-(\ref{section2})) to obtain thermodynamic metrics in a given ensemble, once the first law satisfied at thermodynamic equilibrium is known. We particularly focus on entropic metrics which are generated from the derivatives of the entropy or the free entropy and highlight some key features of such metrics and their relationship with energy metrics (those which are generated from the derivatives of energy functions). The geometry described by entropic metrics is explored in some detail, particularly in the context of black hole chemistry. \\

The paper is organized as follows. In section-(\ref{section2}), we start by setting up our notation and summarize some basic aspects of the contact geometry approach to thermodynamic phase spaces followed by the notion of Hessian metrics defined on spaces of equilibrium states \cite{con1,con2,con2.5,con3,con4}. Then, in section-(\ref{section3}), we discuss some key ideas on thermodynamic Hessian metrics and their reparametrizations. The issue of ensemble non-equivalence is analyzed very carefully. For two-dimensional spaces of thermodynamic equilibrium states, we follow closely the earlier analysis in \cite{bravettiensemble} (see also \cite{Nettel,repar}) and study the (Ruppeiner) thermodynamic curvatures in two different ensembles contrasting their behavior. The possible sources of singularities of the Ruppeiner metric are identified in the two ensembles related by a (partial) Legendre transform. As a model hydrostatic system, we consider the van der Waals fluid and discuss its thermodynamic geometry. In section-(\ref{bh1}), we apply the ideas developed earlier to explore the thermodynamic geometry of black holes in AdS spacetimes in the extended thermodynamics framework. This section contains two subsections, i.e. (\ref{bulk1}) and (\ref{holosec}), where the thermodynamic geometries of the bulk and the boundary (via the gauge/gravity duality) settings are discussed respectively. We end with comments and a summary of the paper in the concluding section-(\ref{section5}).

\section{Contact and metric structures on thermodynamic phase spaces}\label{section2}

In this section, we shall very briefly review some basic aspects of the geometry of thermodynamics. The reader is referred to \cite{Geiges,Arnold,S,CM1,CM2} for the details. Thermodynamic phase spaces assume the structure of a contact manifold, i.e. a \((2n+1)\)-dimensional smooth manifold \(\mathcal{M}\) together with a one form \(\eta\) such that
\begin{equation}\label{contactdef}
  \eta \wedge (d\eta)^n \neq 0.
\end{equation} Clearly, \(\eta \wedge (d\eta)^n\) is a volume form on \(\mathcal{M}\). The kernel of the one form \(\eta\) defines a hyperplane distribution. The condition given in eqn (\ref{contactdef}) is then equivalent to saying that this hyperplane distribution is completely non-integrable in the Frobenius sense or in simpler words the hyperplanes are extremely twisted. This one form \(\eta\) shall be called the contact form. Further, associated with the contact form, there exists a globally defined and unique vector field \(\xi\) known as the Reeb vector field defined through the relations:
\begin{equation}
  \eta(\xi) = 1, \hspace{3mm} d\eta(\xi,.) = 0.
\end{equation}
In other words, the vector field \(\xi\) can be understood to be dual to the one form field \(\eta\). Analogous to the one on symplectic manifolds, there exists a Darboux theorem on contact manifolds which states that on any local patch on a contact manifold \((\mathcal{M},\eta)\), it is always possible to define (Darboux) coordinates \((s,q^i,p_i)\) such that
\begin{equation}\label{local}
  \eta = ds - p_i dq^i, \hspace{3mm} \xi = \frac{\partial}{\partial s}.
\end{equation}
There exist a very special class of submanifolds of a contact manifold \((\mathcal{M},\eta)\) which are of interest especially from the perspective of thermodynamics. They are the integral submanifolds of maximum dimension such that \(\eta=0\) when restricted to the submanifold. In other words, if \(q^i\) and \(p_i\) are to be treated as conjugate variables, it is easy to see that such a submanifold cannot contain a conjugate pair and hence would correspond to the familiar notion of a configuration space from classical mechanics. For a particular Legendre submanifold \(L\) having coordinates \((q^i,p_j)\) where \(i \in I, j \in J\) with \(I\) and \(J\) being a disjoint partition of the index set \(\{1,2,....,n\}\), the local structure is always given as
\begin{equation}\label{Llocal}
  p_i = \frac{\partial F}{\partial q^i}, \hspace{3mm} q^j = - \frac{\partial F}{\partial p_j}, \hspace{3mm} s = F - p_j \frac{\partial F}{\partial p_j}.
\end{equation} In this context \(F=F(q^i,p_j)\) is known as the generator of \(L\) and it should be clear that all Legendre submanifolds are \(n\)-dimensional. It was shown long back that any contact manifold can be associated with a Riemannian metric structure which satisfies some compatibility conditions with the contact form. The reader is referred to the works ~\cite{Sasaki,Hatakeyama,SasakiHatakeyama} for details on compatible metric structures on contact manifolds. The metric is a bilinear, symmetric as well as non-degenerate structure. It can be verified that the generic choice due to Mrugala \cite{Mmetricpaper}: \(G = \eta^2 + dq^i \otimes dp_i\) satisfies all these three basic requirements and also the compatibility condition presented in \cite{Sasaki}. Since for an arbitrary Legendre submanifold \(L\), one has \(\eta|_L=0\) by definition, therefore restricting \(G\) to \(L\) gives the local expression from eqns (\ref{Llocal}):
\begin{equation}\label{GL}
  G|_L = dq^i \otimes dp_i|_L = \frac{\partial^2 F}{\partial q^j \partial q^{j'}} dq^j \otimes dq^{j'} - \frac{\partial^2 F}{\partial p_j \partial p_{j'}} dp_j \otimes dp_{j'}.
\end{equation}
The metric on a Legendre submanifold \(L\) is therefore defined from the Hessian of the generator of \(L\). Such a metric will be called a Hessian metric on \(L\). There are of course other ways of defining a symmetric, bilinear and non-degenerate metric structure on a contact manifold but that would not be of interest to us in the present work.\\

With this background, we can now make connection with thermodynamics (also see \cite{con1,con2,con3,con4}). We start by recalling the first law of thermodynamics for a hydrostatic \((P,V,T)\) system described by the microcanonical ensemble:
\begin{equation}\label{firstlaw}
  dU = TdS - PdV.
\end{equation}
A direct comparison between the first of eqns (\ref{local}) and eqn (\ref{firstlaw}) leads to the immediate identification that the thermodynamic variables are local coordinates on a 5-dimensional contact manifold. Explicitly, one identifies \(s = U\) while \((q^1,q^2) = (S,V)\) and \((p_1,p_2) = (T,-P)\). Further, eqn (\ref{firstlaw}) which holds at equilibrium implies that the system's state is represented by a point on a Legendre submanifold of the thermodynamic phase space. Such a Legendre submanifold has the following local structure [eqn (\ref{Llocal})]:
\begin{equation}
s = U, \hspace{3mm} T = \frac{\partial U}{\partial S}, \hspace{3mm} -P = \frac{\partial U}{\partial V}.
\end{equation} Legendre submanifolds therefore represent spaces of thermodynamic equilibrium states in the sense that each point on the Legendre submanifold represents an equilibrium state of the system. Thus, even though apriori all the coordinates of the thermodynamic phase space are independent, thermodynamic equilibrium or equivalently the first law puts an on-shell condition such that the system lives on a Legendre submanifold with just \(n\) independent coordinates while the other \(n\) are derived by taking derivatives of the thermodynamic potential (the generator) with respect to the independent thermodynamic variables. A thermodynamic system is therefore a triplet \((\mathcal{M},\eta,L)\) where \((\mathcal{M},\eta)\) is a contact manifold and \(L\) is the Legendre submanifold representing the system. It also means that the spaces of thermodynamic equilibrium states are equipped with the notion of a thermodynamic metric which is the Hessian of the relevant thermodynamic potential. In this paper, we are interested in such thermodynamic metrics. The contact geometry approach to thermodynamics naturally leads to a Hamiltonian framework for the latter which for black holes has been discussed in \cite{Rajeev,contactBH} (see also \cite{Bal}). 

\section{Two ensembles related by a Legendre transform}\label{section3}

We shall begin by analyzing two generic ensembles which in the thermodynamic limit are related by a Legendre transform (see also \cite{bravettiensemble} and references therein). Let us say that at equilibrium, the entropy can be expressed as \(S = S(x^i)\) with \(i=1,....,n\) where \(x^i\) are suitable state variables characterzing the system's equilibrium state. For the sake of simplicity, we take the case with \(n=2\) so that we have, \(S = S(E,X)\) where \(E\) is the energy function (for example, internal energy \(U\)) and \(X\) can be a suitable thermodynamic variable.

\smallskip

\subsection{Ensemble \(\mathcal{A}\): \(X\) is fixed by the boundary}\label{e1}
In the thermodynamic limit, we consider the first law with \(E = U\):
\begin{equation}\label{fl}
  dU = TdS + YdX
\end{equation} where \(Y = (\partial U/\partial X)_S\) is the variable conjugate to \(X\). For a hydrostatic system where \(X = V\) (imposed by boundary conditions), one has \(Y = -P\). On the other hand, for a magnetic system one has \(X = \eta \) (the magnetization) and \(Y = h\) (magnetic intensity). Comparison with the first of eqns (\ref{local}) leads us to the identification that \(s = U\) and \((q^1,q^2) = (S,X)\) whereas \((p_1,p_2) = (T,Y)\). The condition, \(dU - TdS - YdX = 0\) defines the space of equilibrium states on which it is most natural to choose \(S\) and \(X\) as the independent coordinates whereas, \(T\) and \(Y\) are defined on-shell as derivatives of the generator function \(U\) with respect to the independent ones. The thermodynamic metric [eqn (\ref{GL})] is then (in our notation, \((dx)^2 = dx \otimes dx\))
\begin{eqnarray}
  dl^2 &=& \frac{\partial^2 U}{\partial S^2} (dS)^2 + 2\frac{\partial^2 U}{\partial S \partial X} dS \otimes dX + \frac{\partial^2 U}{\partial X^2} (dX)^2 \nonumber \\
&=& \frac{T}{C_X} (dS)^2 + 2\bigg(\frac{\partial T}{ \partial X}\bigg)_S dS \otimes dX + \bigg(\frac{\partial Y}{\partial X}\bigg)_S (dX)^2. \nonumber \\
\label{123456}
\end{eqnarray}
This is known as the Weinhold metric \cite{Weinhold}. Note that here \(C_X\) is the the specific heat at constant \(X\). Noting that the function \(U = U(S,X)\) is obtained by inverting the relation \(S = S(U,X)\) in favour of \(U\), then since \(S(U,X)\) is a concave function, \(U(S,X)\) is convex ensuring that the metric given above is positive. This happens because of positivity of temperature, which implies that entropy is a monotonically increasing function of \(U\) \footnote{In eqn (\ref{lineelementgeneric1}), one finds that \(dl^2\) is related to \(dl_R^2\) [eqn (\ref{dlR})] as \(T dl_R^2 = dl^2\) meaning that since \(dl_R^2\) is positive definite, so is \(dl^2\) for \(T > 0\).}. Now, since \(S\) and \(X\) are the independent thermodynamic coordinates on the 2-dimensional space of equilibrium states, such that \(U = U(S,X)\), one has \(T = T(S,X) = \partial_S U (S,X)\) and \(Y = Y(S,X) = \partial_X U (S,X)\). These are the equations of state. Using this, eqn (\ref{123456}) is equivalent to
\begin{equation}\label{naturalmetriccanonical}
  dl^2 = dS \otimes dT + d Y \otimes dX.
\end{equation} This is also easily obtained from eqn (\ref{GL}):
\begin{equation}
  dl^2 = dq^i \otimes dp_i = dq^1 \otimes dp_1 + dq^2 \otimes dp_2
\end{equation} with \(q^1 = S, p_1 = T, q^2 = X, p_2 = Y\). Eqn (\ref{naturalmetriccanonical}) is the line element of the natural metric on the Legendre submanifold \(L_{X}\) representing the system described by this ensemble. \\

Now, because of the equations of state (the on-shell relations between thermodynamic quantities such that only two of them are independent), one can write \(T = T(S,X)\) and \(Y = Y(S,X)\). These can in principle be inverted to obtain \(S=S(T,Y)\) and \(X=X(T,Y)\). One may also obtain \(S=S(T,X)\) and \(Y=Y(T,X)\) or even \(T=T(S,X)\) and \(Y=Y(S,X)\). This means that by suitably inverting the equations of state on the space of equilibrium states \(L_X\), we can pick any two among \(S,T,X\) or \(Y\) to be independent and re-express our metric [eqn (\ref{naturalmetriccanonical})] in four different ways. One is of course eqn (\ref{123456}). The other three are
\begin{eqnarray}
  dl^2 &=& \frac{C_X}{T} (dT)^2 + \bigg(\frac{\partial Y}{\partial X}\bigg)_T(dX)^2, \label{2}\\
  dl^2 &=& \frac{T}{C_Y} (dS)^2 + \bigg(\frac{\partial X}{\partial Y}\bigg)_S (dY)^2, \label{3} \\
  dl^2 &=& \frac{C_Y}{T} (dT)^2 + \bigg(\frac{\partial X}{\partial Y}\bigg)_T (dY)^2  \nonumber \\
  &+& 2\bigg(\frac{\partial X}{\partial T}\bigg)_Y dT \otimes dY. \label{4}
\end{eqnarray} Here, \(C_X\) and \(C_Y\) are the specific heats at constant \(X\) and \(Y\) respectively. It should be specially emphasized that we have not performed any Legendre transformation in deriving these line elements. They are simply eqn (\ref{naturalmetriccanonical}) in different coordinate parameterizations. One goes from one set of independent coordinates to another by exploiting the equations of state while still being on the Legendre submanifold \(L_X\). All these line elements therefore, represent the same length on \(L_X\) but expressed in different fluctuation coordinates. This is possible because the fluctuations in the natural coordinates \((S,X)\) are related to those of the dependent coordinates \((T,Y)\) via the equations of state. Therefore, one expects that the Ricci scalars associated with the line elements given in eqns (\ref{123456}), (\ref{2}), (\ref{3}) and (\ref{4}) are all equivalent to each other. For example, one can compute the scalar curvature on the \((S,X)\) plane [eqn (\ref{123456})] and then using the equations of state re-express it as a function of say, \(T\) and \(Y\). It then means that the curvature scalar so obtained would be the same as that directly calculated using the line element given in eqn (\ref{4}).\\

Since there is a natural first law associated with a given ensemble, this introduces a set of natural coordinates on the Legendre submanifold or the space of thermodynamic equilibrium states on which the system of interest is described. For example, for the present case the natural coordinates are \(S\) and \(X\) although as we saw, by using the on-shell equations of state \(T\) and/or \(Y\) could be made independent on \(L_X\). The choice of natural coordinates does not depend on the specific functional form of the thermodynamic potential (in this case, the internal energy \(U=U(S,X)\)). In an arbitrary case with \(n\) independent variables, one has \(S = S(E,X^j)\) where \(j = 1,2,....,n-1\). One can therefore write \(E = E(S,X^j)\) giving the first law:
\begin{equation}
  dE = TdS + \sum_{j=1}^{n-1} Y_j dX^j.
\end{equation} This sets the natural coordinates to \(\{S,X^j\}\) and the Legendre submanifold describing the system is \(n\)-dimensional. Keeping in mind that the exact form of \(E\) is not relevant here (as long as it is well behaved), one may assert that a given statistical ensemble describes a family of Legendre submanifolds in the thermodynamic limit. The choice of natural coordinates is specified by the ensemble of interest.

\smallskip

\subsection{Ensemble \(\mathcal{B}\): There is a reservoir for \(X\)}\label{e2}
Let us consider the case where the system is in contact with a reservoir for the variable \(X\). For a hydrostatic system, with our usual identification that \(Y\) is the pressure, the bath is a barostat with which the system can exchange its volume. If on the other hand, the system was a magnetic system with \(Y\) being the magnetic intensity, one can think about the system attaining thermodynamic equilibrium in the presence of a constant external field. In the present case, the first law is given by
\begin{equation}\label{fl2}
  dE = TdS - X dY
\end{equation} for some energy function \(E=E(S,Y)\). The first laws given in eqns (\ref{fl}) and (\ref{fl2}) can be related by the Legendre transformation, \(E(S,Y) = U(S,X) - XY\) provided it exists. For a usual hydrostatic system, \(E(S,P) = U(S,V) + PV := H(S,P)\) which is the enthalpy whereas for a magnetic system, \(E(S,h) = U(S,\eta) - \eta h\). \\

Inspecting eqn (\ref{fl2}), we arrive at the following identifications: \((q^1,q^2) = (S, Y)\) and \((p_1,p_2) = (T,-X)\) on the space of equilibrium states (say) \(L_Y\). The thermodynamic length [eqn (\ref{GL})] is then
\begin{equation}\label{alternatenaturalmetric}
  dl^2 = dS \otimes dT - dY \otimes dX
\end{equation} or in the natural coordinates,
\begin{eqnarray}
  dl^2 &=& \frac{\partial^2 E}{\partial S^2} (dS)^2 + 2\frac{\partial^2 E}{\partial S \partial Y} dS \otimes dY + \frac{\partial^2 E}{\partial Y^2} (dY)^2 \nonumber \\
&=& \frac{T}{C_Y} (dS)^2 + 2\bigg(\frac{\partial T}{ \partial Y}\bigg)_S dS \otimes dY - \bigg(\frac{\partial X}{\partial Y}\bigg)_S (dY)^2. \nonumber \\
\label{45}
\end{eqnarray}
Clearly, the thermodynamic lengths on Legendre submanifolds \(L_X\) and \(L_Y\) given respectively in eqns (\ref{naturalmetriccanonical}) and (\ref{alternatenaturalmetric}) are not the same. It can be shown \cite{con4} that two Legendre submanifolds are diffeomorphic to each other if the Legendre transformation connecting them is regular. Even then, the thermodynamic lengths for two ensembles do not coincide. In other words, in the thermodynamic limit where the ensembles become equivalent (up to Legendre transformations), the lengths are not! We strongly emphasize on the fact that this non-equivalence has nothing to do with the microscopic description of a particular system. It is well appreciated that the presence of long ranged interactions may render different ensembles inequivalent to each other \cite{nonequi1,nonequi2} (see also \cite{physrep}). However, the non-equivalence which is being discussed here follows from the basic structure of the thermodynamic phase space and shall continue to be there even when there are no long range interactions between the microscopic degrees of freedom. Thus, non-equivalence in the present context shall refer to the fact that some of the geometrical properties of the two Legendre submanifolds representing the same system but in two different ensembles are not the same. It can be intuitively understood on physical grounds by noting that although one is finally working in the thermodynamic limit, the Hessian metrics are all derived generically based on thermodynamic fluctuations which are not equivalent in different ensembles. As it is clear, the two distinct ensembles are associated with different system-boundary conditions. For example, in ensemble \(\mathcal{A}\), the variable \(X\) is held fixed by the boundary of the system whereas in ensemble \(\mathcal{B}\), the system is in contact with a bath with constant intensive parameter \(Y\). Thus, the fluctuation properties of the thermodynamic system are in general different in the two different ensembles. As it turns out, although in the thermodynamic limit, the behavior of the system consistently agrees in both the ensembles, the thermodynamic lengths which are derived from fluctuation properties within each ensembles are still not equivalent. In other words, the process of taking the thermodynamic limit does not erase the fluctuation properties captured by the thermodynamic lengths.\\

Now for the present case, it is possible to re-express the length [eqn (\ref{alternatenaturalmetric})] in different coordinate parameterizations. One of them is eqn (\ref{45}). The other three are
\begin{eqnarray}
  dl^2 &=& \frac{C_Y}{T} (dT)^2 - \bigg(\frac{\partial X}{\partial Y}\bigg)_T(dY)^2, \label{5}\\
  dl^2 &=& \frac{T}{C_X} (dS)^2 - \bigg(\frac{\partial Y}{\partial X}\bigg)_S (dX)^2, \label{6} \\
  dl^2 &=& \frac{C_X}{T} (dT)^2 - \bigg(\frac{\partial Y}{\partial X}\bigg)_T (dX)^2  \nonumber \\
  &-& 2\bigg(\frac{\partial Y}{\partial T}\bigg)_X dT \otimes dX. \label{7}
\end{eqnarray}
It turns out that the Ricci scalars of the line elements given in eqns (\ref{45}), (\ref{5}), (\ref{6}) and (\ref{7}) are equivalent to one another. However, the line elements with the same fluctuation coordinates (say \((T,X)\)) are not equivalent in the two ensembles. Therefore, to summarize, the Ricci scalars associated with thermodynamic metrics corresponding to different ensembles (hence, different families of Legendre submanifolds) are in general inequivalent.

\smallskip

\subsection{Entropic metrics}\label{entropicmetricssubsection}
So far we saw that it is possible to construct various thermodynamic metrics by taking Hessians of different thermodynamic potentials. Typically, such potentials are the energy functions of the system such as the internal energy or the enthalpy. However, it is often physically more intuitive to consider entropic potentials (those with dimensions of entropy) in the construction of such metrics. Among them the Ruppeiner metric is special because it is directly linked with the probability of fluctuations rendering a physical meaning to the length between two thermodynamic states. Furthermore, its Ricci scalar, i.e. the Ruppeiner curvature or the thermodynamic curvature bears a nice physical interpretation as was pointed out in the introduction. \\

As it turns out, eqn (\ref{fl}) can be re-written as
\begin{equation}\label{microcanonicalfirstlawentropy}
	dS = \frac{dU}{T} - \frac{Y}{T} dX.
\end{equation} This is clearly a microcanonical description where the entropy is given by the Boltzmann formula \(S = \ln \Omega (U,X)\). From the point of view of statistical mechanics, this is a more fundamental form of the first law as compared to eqn (\ref{fl}) because all the equilibrium properties including the specific heats and susceptibilities can be computed from the knowledge of \(S\) derived from microscopic details (via \(\Omega\)). The Ruppeiner metric is then defined as the negative Hessian of the entropy or equivalently, the Hessian of the negative entropy. In order to derive an expression for the Ruppeiner metric in ensemble \(\mathcal{A}\), let us start out with the generic expression for the Ruppeiner line element, \(dl_R^2 = -g_{i j} dx^i \otimes dx^j\) with \(g_{i j} = \partial_i \partial_j S \). Writing out \(dz_i = g_{i j} dx^j\) one finds
\begin{equation}\label{1}
  dl_R^2 = - dz_i \otimes dx^i.
\end{equation}
Now, since
\(dz_i = g_{i j} dx^j\), we must have
\begin{equation}
  z_i = \frac{\partial S}{\partial x^i}.
\end{equation}
From the first law given in eqn (\ref{fl}), one can write
\begin{equation}\label{firstlawcontact}
  dS - \frac{dU}{T} + \frac{Y dX}{T} = 0
\end{equation}
which means that \(z_1 = 1/T\) and \(z_2 = - Y/T\) whereas \(x^1 = U\) and \(x^2 = X\). With these identifications,
\begin{equation}
  dz_1 = -\frac{dT}{T^2}, \hspace{2mm} dz_2 = \frac{Y dT}{T^2} - \frac{dY}{T}.
\end{equation} The line element given in eqn (\ref{1}) can now be expressed as
\begin{equation}
  dl_R^2 = - \bigg(-\frac{dT}{T^2}\bigg) \otimes dU - \bigg(\frac{Y}{T^2}dT - \frac{dY}{T}\bigg) \otimes dX
\end{equation} which from the first law reduces to
\begin{equation}\label{lineelementgeneric1}
  dl_R^2 = \frac{1}{T} (dS \otimes dT + d X \otimes dY).
\end{equation}
We can now turn to ensemble \(\mathcal{B}\) where \(Y\) is fixed by an external bath. In this case the Ruppeiner line element is
\begin{equation}\label{ruppensembleB}
  dl^2_R = \frac{1}{T} (dS \otimes dT - d Y \otimes dX)
\end{equation} where the entropy of the system is of the generic form \(S = S(E,Y)\). This is different from the microcanonical or \((V,U)\)-description. The first law in terms of the entropy can be re-written as
\begin{equation}
	dS = \frac{dE}{T} + \frac{X}{T} dY
\end{equation} which is equivalent to rearranging eqn (\ref{microcanonicalfirstlawentropy}) and defining \(E = U - YX\). Since \(X\) is an extensive variable, its conjugate, \(Y\) is intensive and consequently \(S\) has been expressed as a function of an intensive and an extensive variable as opposed to the microcanonical description where it is a function of \(U\) and \(X\), both being extensive. It therefore follows that in ensemble \(\mathcal{B}\), the entropy is convex in argument \(Y\) while still being concave in \(E\). Nevertheless, its Hessian is still negative making it a concave function overall. This ensures that the Ruppeiner metric defined as the negative Hessian of the entropy is positive. The energy metric defined as the Hessian of \(E\) discussed in the previous subsection, being conformally related to the Ruppeiner metric [eqn (\ref{ruppensembleB})] is also positive for \(T > 0\). If the system of interest is a hydrostatic system, then such a description would correspond to the \((H,P)\)-ensemble where the pressure \(P\) and the enthalpy \(H\) are held fixed at the boundary.\\

As it turns out, the curvature scalar associated with the Ruppeiner metric has a peculiar behavior close to critical points or even the points at which the system gets strongly correlated \cite{Rup2}. Let us examine the case of a system described by the ensemble \(\mathcal{A}\) with two independent thermodynamic coordinates. For such a case, the metric has the general form: \(dl_R^2 = g_{11} (dx^1)^2 + g_{12} dx^1 \otimes dx^2 + g_{21} dx^2 \otimes dx^1 + g_{22} (dx^2)^2\) where \(g_{12}= g_{21}\). The Ricci scalar corresponding to the geometry described by the metric can be obtained to be \cite{R}
\begin{eqnarray}\label{RR}
    R = - \frac{1}{\sqrt{g}}\bigg[\frac{\partial}{\partial x^1}\bigg(\frac{g_{12}}{g_{11}\sqrt{g}}\frac{\partial g_{11}}{\partial x^2} - \frac{1}{\sqrt{g}}\frac{\partial g_{22}}{\partial x^1}\bigg)  \nonumber \\
 \frac{\partial}{\partial x^2}\bigg(\frac{2}{\sqrt{g}}\frac{\partial g_{12}}{\partial x^1} - \frac{1}{\sqrt{g}}\frac{\partial  g_{11}}{\partial x^2} - \frac{g_{12}}{g_{11}\sqrt{g}}\frac{\partial g_{11}}{\partial x^1}\bigg) \bigg]
\end{eqnarray}
where \(g\) is the determinant of the metric tensor. This means that calculations are a lot simpler if the metric is diagonal. For the sake of simplicity, let us consider the ensemble \(\mathcal{A}\) discussed in a preceding subsection and pick up the Ruppeiner element \(dl_R^2\) obtained by dividing the line element given in eqn (\ref{2}) by a factor of \(T\):
\begin{equation}\label{RRR}
  dl_R^2 = \frac{C_X}{T^2} (dT)^2 + \frac{1}{T}\bigg(\frac{\partial Y}{\partial X}\bigg)_T (dX)^2.
\end{equation}
Clearly, one finds that the metric is singular if \(C_X = 0\) or \((\partial Y/\partial X)_T = 0\). Taking the Ruppeiner line element obtained by dividing eqn (\ref{3}) by a factor of \(T\), it also follows that the curvature scalar say \(R_U\) in ensemble \(\mathcal{A}\) can also diverge as \(C_Y \rightarrow \infty\). Blowing up of the specific heat is reminiscent of a critical point which indicates that \(R_U\) may blow up at the critical point. For a hydrostatic system where \(Y = -P\) and \(X = V\), the situation corresponds to blowing up of \(C_P\) at the critical point.
\\

Next, let us consider the ensemble \(\mathcal{B}\) consisting of a system in contact with a bath for \(X\). It is not hard to convince oneself that the thermodynamic curvature obtained in this case, say \(R_E\) does not coincide with \(R_U\). The metrics and hence their Ricci scalars simply do not match. However, the curvature scalar \(R_E\) diverges as \(C_X \rightarrow \infty\) and \(C_Y = 0\) suggesting a complimentary behavior to \(R_U\) as far as divergences are concerned. For a general case where there are two fluctuation variables, the correspondence between the zeroes and divergences of the specific heats with the singularities of the thermodynamic lengths (which could possibly lead to divergences of the curvature scalars) has been summarized in table-(1).
\begin{table*}[t]
\caption{Possible sources of singularities of the Ruppeiner metric.\\}
\centering
\begin{tabular}{| c | c | c | c |}
\hline
{\bf Ensemble}  & {\bf Possible sources of singularities of the Ruppeiner metric}    \\ \hline
\((U,X)\)           &  (a) Divergences of \(C_Y\) (b) Zeroes of \(C_X\) (c) At \((\partial Y/\partial X)_T = 0\) (d) At \((\partial X/\partial Y)_S = 0\)   \\
\((E,Y)\)                          &  (a) Divergences of \(C_X\) (b) Zeroes of \(C_Y\) (c) At \((\partial X/\partial Y)_T = 0\) (d) At \((\partial Y/\partial X)_S = 0\)  \\ \hline
\end{tabular}
\end{table*}

\smallskip

\subsection{($U,V)$ and $(H,P)$-ensembles}
In this subsection, we shall make our assertions concrete by considering the van der Waals model, which exhibits features of the liquid-gas phase transition. Then, ensemble \(\mathcal{A}\) discussed in subsection-(\ref{e1}) is the familiar microcanonical or \((U,V)\)-ensemble where symbols have their usual meaning. On the other hand, ensemble \(\mathcal{B}\) discussed in subsection-(\ref{e2}) then corresponds to the isoenthalpic-isobaric or \((H,P)\)-ensemble \cite{NPH} with energy function \(H = U + PV\). For a hydrostatic system, our notations and conventions are as follows: The number of particles \(N\) is kept fixed and is not allowed to fluctuate. Thus \(N\) is merely a fixed parameter setting the system's size. For convenience, we put \(X = v\) (rather than \(V\)) and we have \(Y = -P\) where \(v=V/N\) is now the specific volume of the fluid such that the ideal gas equation reads \(Pv = T\). \\

\begin{figure*}
 \includegraphics[width=3.4in]{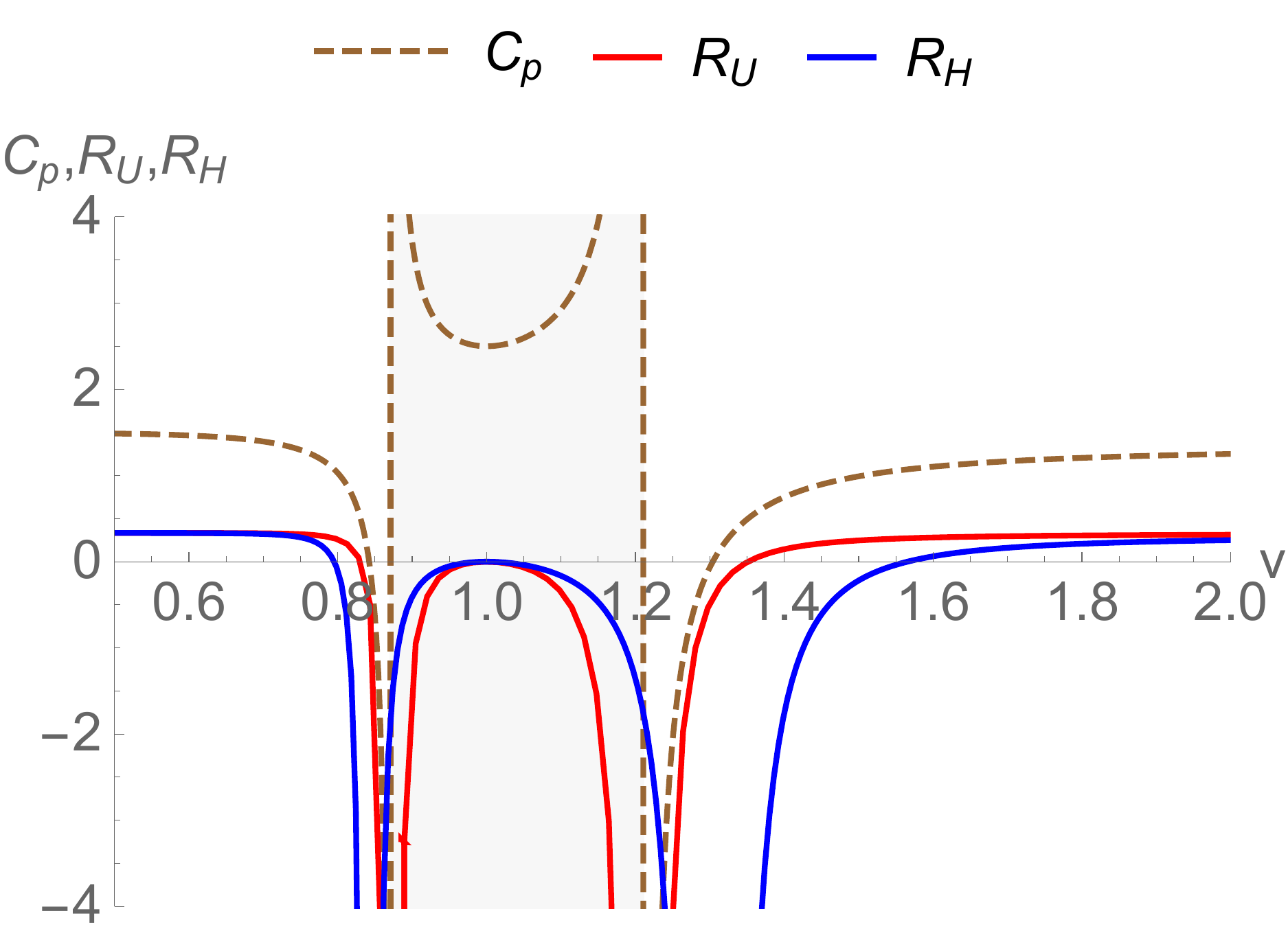}
  \includegraphics[width=3.4in]{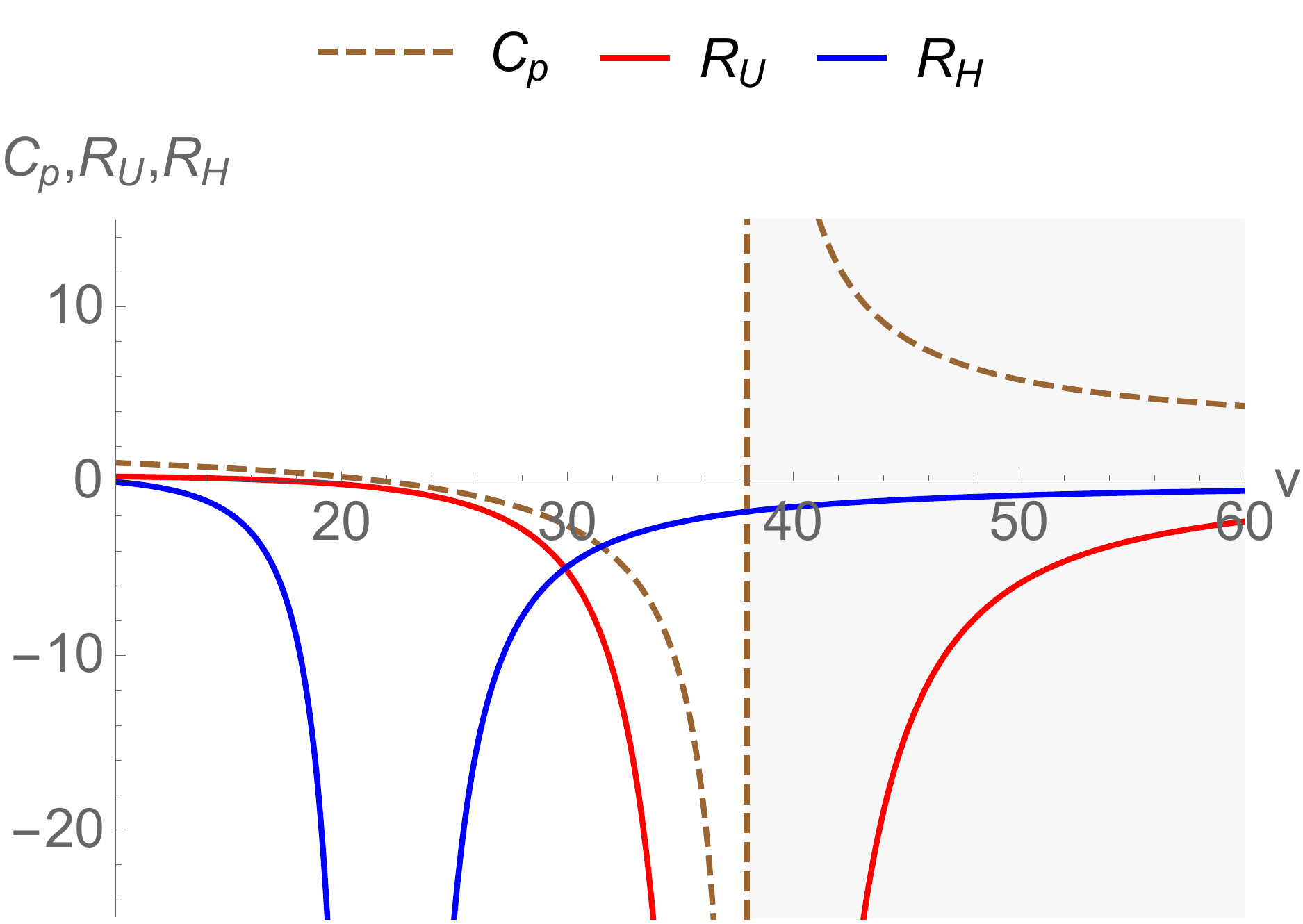}
\caption{Thermodynamic curvatures \(R_U\), \(R_H\) and specific heat \(C_P\) for the van der Waals fluid plotted versus specific volume with $a=2, b=1$ below the critical temperature. The region(s) with negative bulk modulus are shaded in grey.}
\label{RU_R_H_Cp_vdw}
\end{figure*}

The van der Waals (vdW) fluid is a prototypical example of a model fluid exhibiting features of the liquid-gas phase transition. The attractive interactions among the fluid molecules can be summarized by the van der Waals potential \(V(r) = - k/r^6\) for some constant \(k > 0\) acting between pairs of molecules. Furthermore, the molecules are modelled as impenetrable hard spheres and therefore, if \(\sigma\) be the distance at which two molecules touch each other, the potential is infinite. Such a description of intermolecular interactions is mean field. The fluid is described by the equation of state:
\begin{equation}
P=\frac{T}{v-b}-\frac{a}{v^2}.
\end{equation}
Taking \(C_v = 3/2\), one has the following expression for \(C_P\):
\begin{equation}
C_P=\frac{T v^3}{T v^3-2 a (b-v)^2}+\frac{3}{2}.
\end{equation} Note that putting \(a=b=0\), one has \(C_P - C_v = 1\) as expected from the ideal gas. In the microcanonical ensemble, the thermodynamic curvature reads~(see also \cite{Wei1,Wei2})
\begin{equation}
R_U=-\frac{4 a (b-v)^2 \left(a (b-v)^2-T v^3\right)}{3 \left(T v^3-2 a
   (b-v)^2\right)^2}
\end{equation}
whereas, in the isoenthalpic-isobaric ensemble, the curvature scalar of the Ruppeiner metric reads the following when expressed in \(T\) and \(v\) coordinates:
\begin{equation}
R_H=-\frac{4 a (b-v)^2 \left(3 a (b-v)^2-5 T v^3\right)}{\left(6 a (b-v)^2-5
   T v^3\right)^2}
\end{equation} where the subscript \(H\) signifies that in this case, the energy function \(E\) is equal to the enthalpy. The thermodynamic curvatures \(R_U\) and \(R_H\) have been plotted as a function of \(v\) in figures-(\ref{RU_R_H_Cp_vdw}). 
Clearly, the divergences in \(C_P\) and \(R_U\) exactly coincide as was expected. With \(v > b = 1\), $C_P$ has divergences at $v=1.21,37.9$ and $R_U$ diverges exactly at these points. We mention that the plots are for a temperature below the critical temperature. The curvature scalar \(R_U\) becomes zero at the point \(v=b\) where the specific volume and the co-volume coincide putting a lower cutoff to the physical values \(v\) can take. However, this point lies in the negative bulk modulus range below the critical temperature. We also note that \(R_U\) crosses zero for two values of \(v > b\). However, both of them occur in the region where the isothermal bulk modulus is negative thus falling in a thermodynamically unstable region. For $T=0.1$, $a=2$ and $b=1$, the isothermal bulk modulus is negative (shaded in grey in the plots) between \(v = 1.210\) and \(37.918\)). Such crossings are therefore not considered to be of physical interest. It is then simple to check that the thermodynamic curvature is negative definite over the entire thermodynamically stable region with \(v > b\) and \((\partial P/\partial v)_T \leq 0\). This can empirically be taken to signal the existence of attractive interactions between molecules. It may be shown that near the critical temperature \(T_c\), the specific heat \(C_P\) and thermodynamic curvature \(R_U\) scale as \cite{Wei2}
\begin{equation}
C_P \sim |T - T_c|^{-1}, \hspace{5mm} R_U \sim |T - T_c|^{-2}.
\end{equation} The exponent `2' for the thermodynamic curvature near the critical point has been obtained earlier in other contexts \cite{Wei1,exponent,information,chemical5,unicrit}. \\

As for \(R_H\), the divergences of \(R_H\) do not correspond to those of \(C_P\). As a matter of fact, if \(C_v\) had divergences, one could expect such divergences to coincide with those of the thermodynamic curvature \(R_H\). In the present case where \(C_v\) is a constant it can be clearly seen that the divergences of \(R_H\) correspond to the zeroes of \(C_P\). It should be emphasized that the constancy of one of the specific heats (here \(C_v\)) originates from the specific choice \(X = v\) for the \((U,X)\) ensemble where \(C_v\) is a constant due to the equipartition theorem for the van der Waals fluid. For \(v > b = 1\), $C_P$ has zeroes at $v=21.8$ and $1.3$. $R_H$ diverges at both these points. This is expected from the generic structure of the line elements presented earlier. Let us also note that \(R_H\) consistently goes to zero at \(v = b\). Finally, we point out that the other crossings of \(R_H\) fall into the region of negative isothermal bulk modulus (shaded in grey) and are therefore discarded. Thus, \(R_H\) is negative over the entire physically interesting region possibly signifying the attractive nature of van der Waals interactions between the molecules. Furthermore, let us note that as one takes \(v \rightarrow \infty\), both \(R_U\) and \(R_H\) approach zero. This is the ideal gas limit where the thermodynamic geometry is flat. \\

Some discussion is in order. Although we find that for the van der Waals model, the physically interesting zero of both \(R_U\) and \(R_H\) agree, irrespective of their inequivalence, this is not true in general. For example, if we use the fact that the specific heat at fixed volume is a constant (certainly true for several model systems), then \(R_H\) and \(R_U\) have the following general expressions (written in terms of specific volume):
\begin{widetext}
\begin{equation}
R_H = -\frac{T^2 C_v \partial_{T,v}P^2-2 T^2 C_v \partial_{v}P \partial_{T,T,v}P-C_v \partial_{v}P^2+2 T^3
   \partial_{T}P^2 \partial_{T,T,v}P-2 T^2 \partial_{T}P^2 \partial_{T,v}P+2 T \partial_{T,v}P^2 \partial_{v}P^2}{2 \left(T
   \partial_{T}P^2-C_v \partial_{v}P\right){}^2}
\end{equation} and,
\begin{equation}
R_U= \frac{T^2 \left(-\partial_{T,v}P^2\right)+2 T^2 \partial_{v}P \partial_{T,T,v}P+\partial_{v}P^2}{2 C_v
   \partial_{v}P^2}.
\end{equation}
\end{widetext}
Now, using the fact that \(\partial_{T,T,v}P=0\) (the equation of state is linear in \(T\)), it is simple to check that both \(R_U = R_H = 0\) when
\begin{equation}
  \bigg(\frac{\partial P}{\partial v}\bigg)_{T = 0} = 0 .
\end{equation}  The above condition, gives one physical solution at which both \(R_U\) and \(R_H\) vanish, both below and above the critical point. Other zero crossings are physically not quite interesting because they lie in the range of negative bulk modulus. Although the above result proving the equivalence of the zero crossing(s) of \(R_U\) and \(R_H\) looks appealing, let us emphasize that it is based on two crucial assumptions about the fluid system. First, we have assumed that \(C_v\) is a constant, independent of \(T\) and \(v\). Although this follows from the equipartition theorem, this is certainly not true for a general fluid with complicated microscopic interactions where the virial coefficients are temperature dependent. The second assumption is that the equation of state is linear in \(T\), i.e. \(\partial_{T,T,v}P=0\). While this is true for the ideal gas and several model fluid systems (such as van der Waals), this is not the case for a general fluid where the virial coefficients do not depend on temperature linearly. Thus, the zero crossing behavior of \(R_U\) and \(R_H\) are indeed not equivalent in the general case. Nevertheless, they do follow some general trends as far as their divergences are concerned.

\section{Black holes in AdS spacetimes} \label{bh1}
We shall consider black holes in AdS spacetimes. In the extended thermodynamics framework \cite{Kastor:2009wy}, the cosmological constant is treated as thermodynamic pressure via the relation
\begin{equation}
P = - \frac{\Lambda}{8 \pi} \hspace{7mm} {\rm with,} \hspace{7mm} \Lambda = - \frac{(d-1)(d-2)}{2l^2}. 
\end{equation} Here \(d\) is the number of spacetime dimensions. For charged black holes, the first law of thermodynamics takes the following form:
\begin{equation}
dM = TdS + VdP + \Phi dQ
\end{equation} where, \(Q\) is the electric charge (\(U(1)\) charge) of the black hole, and \(\Phi\) is the corresponding potential. The thermodynamic variables satisfy the Smarr relation \cite{Kastor:2009wy,Kubiznak:2012wp,Gunasekaran:2012dq}:
\begin{equation}
(d-3) M = (d-2) TS - 2 PV + (d-3) Q \Phi
\end{equation} which can be obtained via scaling arguments. Thermodynamic geometry of black holes was first studied in \cite{Cai} wherein the BTZ black hole was considered and it was found that the curvature scalar diverges at extremality. This was followed by a series of papers (see for instance \cite{Aman:2003ug,Shen:2005nu,Mirza:2007ev,Quevedo:2007mj}) where the thermodynamic curvature for various black holes were computed and analyzed. It was found that upon suitably choosing the thermodynamic potential, the thermodynamic curvature is divergent along the Davies line \cite{Shen:2005nu} (see also \cite{bravettiensemble}). The most natural choice of energy metric for black holes is defined as the Hessian of the mass \cite{Cai,Aman:2003ug,Shen:2005nu}, i.e. 
\begin{equation}
dl_M^2 = \frac{\partial^2 M}{\partial y^i \partial y^j} dy^i \otimes dy^j
\end{equation} where \(M = M(y_i)\) with \(y_1 = S\) (entropy). One may invert this fundamental relation, to express the entropy as the potential, ie. \(S = S(M, \cdots)\) and subsequently define the Ruppeiner metric \(dl^2_R\) from it. Using arguments identical to those discussed in subsection-(\ref{entropicmetricssubsection}), it follows that \(dl_R^2 = dl_M^2/T\) where, \(T\) is the Hawking temperature. There have been several extensive investigations on thermodynamic geometry of black holes in the literature \cite{Wei1,Wei2,Cai,Aman:2003ug,Shen:2005nu,Sarkarbtz,Mirza:2007ev,Quevedo:2007mj,Banerjee:2010da,Sahay:2010tx,Liu:2010sz,AR,Wei:2015iwa,Mansoori:2016jer,Bhattacharya:2017hfj,Miao:2017fqg,Xu,weiGB,meanfield1,wei4DGB,mansoori4DGB,btz,btz2,meanfield2,Dehyadegari:2020ebz,HosseiniMansoori:2020jrx,unicrit,NaveenaKumara:2020biu,Wei:2020kra,Yerra:2020tzg,chemical1,chemical2,chemical3,chemical4,chemical5}. We remark here that although we shall be focussing on Hessian metrics, there are other metric structures which have been considered for black holes earlier \cite{Quevedo:2007mj,geometrotherm}.\\

In the extended thermodynamics framework, thermodynamic geometry for BTZ black holes (\(d = 3\)) has been studied in \cite{btz,btz2} (also see \cite{Cai,Sarkarbtz} for older studies). It has been found that for the neutral and non-rotating BTZ black hole, the thermodynamic geometry is Ricci flat, empirically indicating towards the absence of net microscopic interactions. However, for black holes with electric charge and/or angular momentum, the geometry is curved with a positive thermodynamic curvature. This may be taken to indicate towards the presence of repulsive microscopic interactions and is consistent with the fact that the BTZ black hole does not admit a phase transition \cite{lower}. For charged and/or rotating BTZ black holes, the scalar \(R_H\) has been found to diverge at the extremal point \cite{btz}, consistent with a much older result \cite{Cai}. The thermodynamic curvature \(R_H\) was obtained for the case of exotic BTZ black holes in \cite{btz}, and it was found that \(R_H\) could be both positive and negative with a zero crossing between the two regimes. The origin of such a crossing is not well understood, partly because exotic BTZ black holes do not admit a fluid-like equation of state. On the other hand, thermodynamic curvatures obtained for black holes in higher dimensions exhibit richer features. Below, we shall consider charged AdS black holes in four dimensions. 

\smallskip

\subsection{Bulk} \label{bulk1}
 The solution to Einstein-Maxwell equations with a negative cosmological constant in four dimensions (\(d=4\)) reads \cite{Kubiznak:2012wp}:
\begin{equation}
ds^2 = - f(r) dt^2 + f(r)^{-1} dr^2 + r^2 d\Omega_2^2,
\end{equation}
\begin{equation}
A = -\frac{q}{r} dt, \hspace{10mm} F = dA
\end{equation}
 where \(d\Omega_2^2\) is the line element on a 2-sphere and, 
 \begin{equation}
 f(r) = 1 - \frac{2M}{r} + \frac{q^2}{r^2} + \frac{r^2}{l^2}. 
 \end{equation} Here, \(M\) is the ADM mass and \(q\) is the \(U(1)\) charge of the spacetime. The event horizon is defined as the largest root of the relation \(f(r_+) = 0\). In terms of \(r_+\), the black hole mass can be expressed as
 \begin{equation}
 M = \frac{r_+}{2} \bigg(1 + \frac{q^2}{r_+^2} + \frac{r_+^2}{l^2}\bigg).
 \end{equation} 
 It should be remarked that here, the mass takes the role of the enthalpy of the system, i.e. \(M := H(S,P,q)\) \cite{Kastor:2009wy}. In terms of thermodynamic variables \(S = \pi r_+^2\) and \(P = 3/8 \pi l^2\), the enthalpy (mass) is given by
\begin{equation}
  H(S,P,q) = \frac {1} {6\sqrt{\pi}} S^{-\frac{1}{2}}
\left(8 P S^2 + 3S+3\pi q^2 \right).
\end{equation} Temperature and thermodynamic volume can be computed by differentiating the enthalpy giving
\begin{equation}
  T = \bigg(\frac{\partial H}{\partial S}\bigg)_P, \hspace{3mm}  V = \bigg(\frac{\partial H}{\partial P}\bigg)_S
\end{equation} which upon elimination of \(S\), gives the equation of state:
 \begin{equation}\label{eqnRNADSBH}
P=\frac{2 (6 \pi )^{2/3} q^2+6 (6 \pi )^{2/3} T V-3 \sqrt[3]{6} V^{2/3}}{36
   \sqrt[3]{\pi } V^{4/3}}.
\end{equation}
This provides an on-shell relationship between \(P\), \(V\) and \(T\) for the charged AdS black hole. Thermodynamic geometry of the system has been studied earlier in both the \((U,V)\)-ensemble \cite{Wei1,Wei2} and the \((H,P)\)-ensemble \cite{meanfield1,AR}. Let us note that the thermodynamic volume turns out to be (see also \cite{Cvetic:2010jb})
\begin{equation}
V = \frac{4}{3} \frac{S^{3/2}}{\sqrt{\pi}}
\end{equation}
which is only a function of entropy (no pressure dependence). Therefore, the specific heat at constant volume \(C_V\) identically vanishes ensuring that \(R_U\) is divergent for all thermodynamic equilibrium states. The authors of \cite{Wei1,Wei2} have suggested a remedy by considering \(C_V\) to be a vanishingly small number (rather than zero), of the order of \(k_B\) and then one may define a normalized curvature as
\begin{equation}
\tilde{R}_U = \lim_{C_V \rightarrow 0^+} C_V R_U
\end{equation} which is finite. As a matter of fact, in black hole chemistry with \(d \geq 3\), \(C_V\) vanishes for all black holes with spherical symmetry and as such the procedure described above works for all such cases. In what follows, we compare and contrast the behavior of the curvature scalars obtained in \((U,V)\) and \((H,P)\)-ensembles \cite{meanfield1,AR}.\\

\begin{figure}[t]
\begin{center}
\includegraphics[width=3.2in]{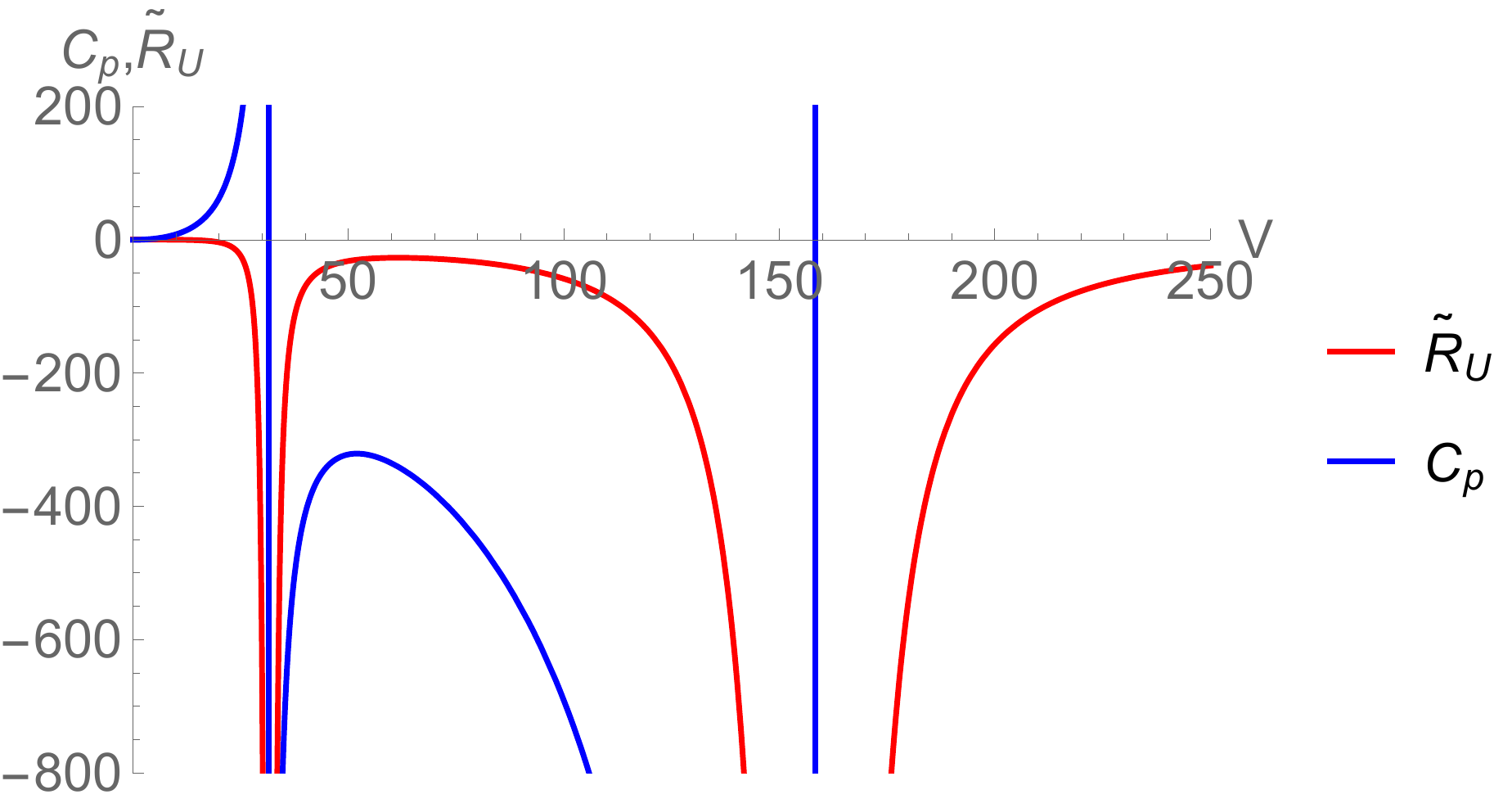}
\caption{Normalized thermodynamic curvature \(\tilde{R}_U\) for the RN-AdS black hole and specific heat \(C_P\) plotted versus thermodynamic volume \(V\) with $q=1, T=0.01$.}
\label{RU_Cp_rnbh}
\end{center}
\end{figure}

\begin{figure}[t]
\begin{center}
\includegraphics[width=3.2in]{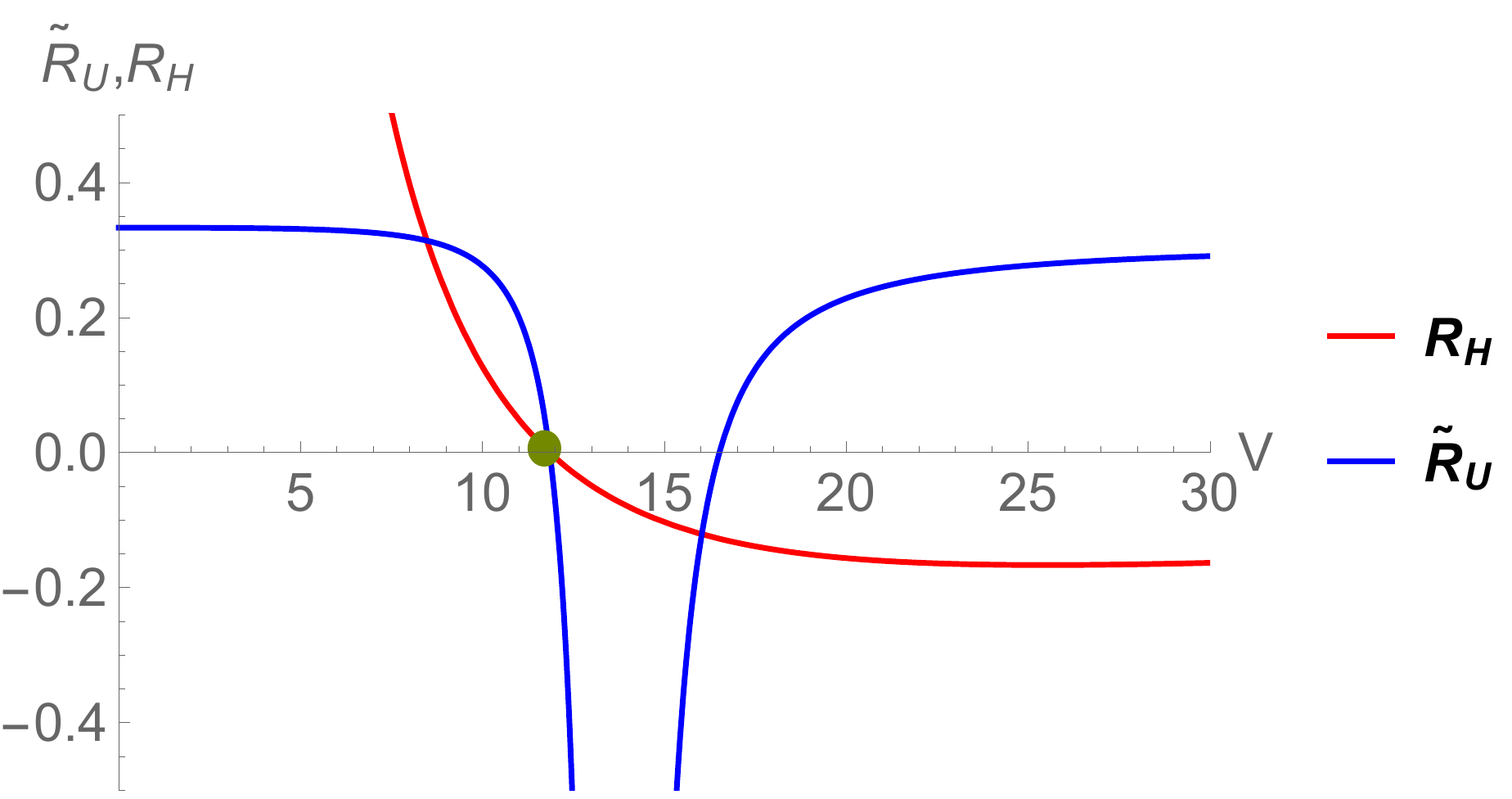}
\caption{Zero crossing of \(\tilde{R}_U\) and \(R_H\) (with \(C_V = 0\)) for the RN-AdS black hole plotted versus thermodynamic volume \(V\) with $q=1, T=0.01$.}
\label{crossing}
\end{center}
\end{figure}

The specific heat \(C_P\) turns out to be
\begin{equation}
C_P=\frac{9 \pi  T V^{5/3}}{\sqrt[3]{6} \pi ^{2/3} \left(4 q^2+3 T V\right)-3 V^{2/3}}
\end{equation} and we have the following expression for \(\tilde{R}_U\):
\begin{equation}\label{Ru}
  \tilde{R}_U= \frac{A_1 \times A_2}{A_3}
\end{equation} where,
\begin{eqnarray*}
  A_1 &=& \left(4 \sqrt[3]{6} \pi ^{2/3} q^2-3 V^{2/3}\right), \nonumber \\
  A_2 &=& \left(2 \sqrt[3]{6} \pi^{2/3} \left(2 q^2+3 T V\right)-3 V^{2/3}\right), \nonumber \\
  A_3 &=& 3 \left(\sqrt[3]{6} \pi ^{2/3}
   \left(4 q^2+3 T V\right)-3 V^{2/3}\right)^2. \nonumber
\end{eqnarray*} \(\tilde{R}_U\) has been plotted together with \(C_P\) in figure-(\ref{RU_Cp_rnbh}).
One clearly sees that the divergences of \(\tilde{R}_U\) coincide with those of \(C_P\). The only zero crossing of the normalized thermodynamic curvature which falls in the region of thermodynamic stability occurs at
\begin{equation}
  4 \sqrt[3]{6} \pi ^{2/3} q^2 = 3 V^{2/3}.
\end{equation} This in terms of the horizon radius \(r_+\) is equivalent to the condition: \(r_+ = \sqrt{2}|q|\). On the other hand, the thermodynamic curvature in the isothermal-isobaric ensemble is regular, i.e. it does not need to be normalized. It reads
\begin{equation}\label{Rh}
  R_H= \frac{B_1 \times B_2}{B_3}
\end{equation} where,\begin{eqnarray*}
  B_1 &=& \left(4 \sqrt[3]{6} \pi ^{2/3} q^2-3 V^{2/3}\right), \nonumber \\
  B_2 &=&  (4
   \sqrt[3]{6} \pi ^{2/3} q^2 C_V + 6 \sqrt[3]{6} \pi ^{2/3} T V C_V \nonumber\\
   &-&3V^{2/3} C_V+18 \pi  T V^{5/3}), \nonumber \\
  B_3 &=& 2 (4 \sqrt[3]{6} \pi
   ^{2/3} q^2 C_V+3 \sqrt[3]{6} \pi ^{2/3} T V C_V \nonumber\\
   &-&3 V^{2/3} C_V+9 \pi \nonumber
    T V^{5/3})^2.
\end{eqnarray*}
This expression coincides with the one obtained earlier in \cite{meanfield1} for the choice $C_V=0$. Since, the limit \(C_V \rightarrow 0^+\) is smooth for \(R_H\), we can very well set it equal to zero without the need of normalizing the thermodynamic curvature unlike the case of \(R_U\). The scalar \(R_H\) diverges as \(V \rightarrow 0\). This corresponds to the limit \(C_P \rightarrow 0\) and is consistent with our expectations. Remarkably, even for the black hole, the two thermodynamic curvatures have identical crossing points within the region of thermodynamic stability (corresponding to the horizon radius \(r_+\) satisfying \(r_+ = \sqrt{2}|q|\)) independent of \(C_V\) \cite{meanfield1}. This is shown in figure-(\ref{crossing}) and can also been seen by noticing that the factors $A_1$ and $B_1$ appearing in eqns (\ref{Ru}) and (\ref{Rh}) are the same, independent of the value (constant) of $C_V$ chosen.
 This is because in the equation of state, i.e. eqn (\ref{eqnRNADSBH}), the pressure \(P\) depends on \(T\) linearly. It should be noted that unlike the van der Waals fluid, where the thermodynamic curvatures were negative definite over the entire physical range, for charged black holes in AdS there is a region in which the thermodynamic curvatures are positive, possibly implying towards the existence of repulsive interactions \cite{Wei1}. However, if we set \(q = 0\), i.e. consider (neutral) Schwarzschild-AdS black holes, the thermodynamic curvatures are negative definite. If one considers the sign of the thermodynamic curvatures to indicate towards the nature of microscopic interactions, at least empirically, then this gives rise to an interesting picture. One may speculate that a charged black hole in AdS is associated with two kinds of microscopic degrees of freedom \cite{AR,meanfield1}. The first type, which are present even in the \(q = 0\) case are associated with attractive interactions whereas, the second type, which are present only when \(q \neq 0\), interact in a repulsive manner. Therefore, Schwarzschild-AdS (\(q = 0\)) black holes have \(\tilde{R}_U, R_H < 0\) whereas, for their electrically charged counterparts, the thermodynamic curvatures can be both positive and negative (even zero) depending on the competition between the two distinct kinds of microscopic degrees of freedom. However, one should bear in mind that these remarks are mere speculations and cannot substitute for independent microscopic computations to describe the statistical mechanics of black holes.  \\

\begin{figure}[t]
\begin{center}
\includegraphics[width=3.2in]{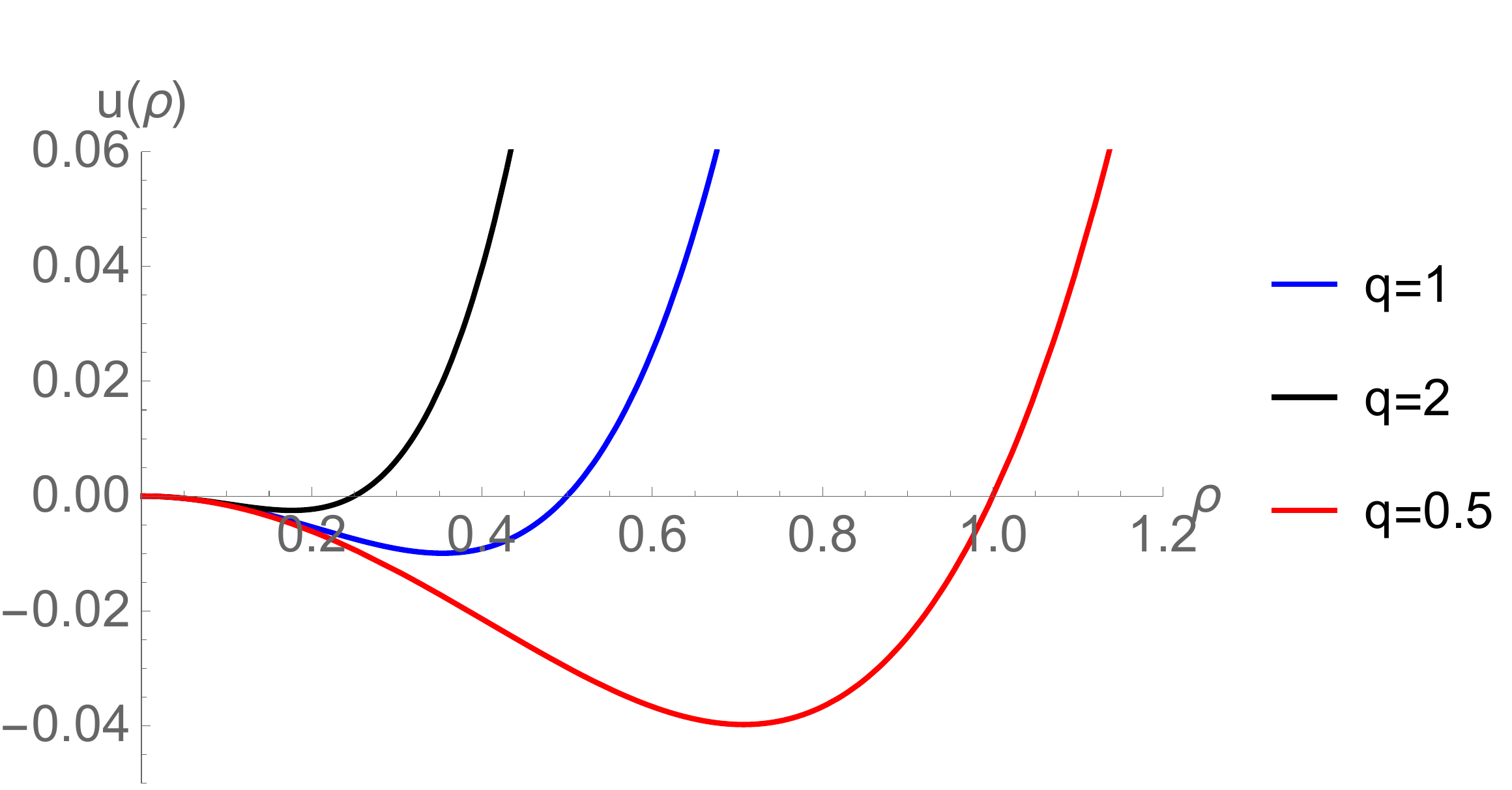}
\caption{Mean field potential \(u(\rho)\) for different values of electric charge.}
\label{potential}
\end{center}
\end{figure}

Interestingly, the existence the zero crossings for black holes may be explained naively as follows. If we define a specific volume \(v = 2 r_+ = 2 (3V/4 \pi)^{1/3}\), then eqn (\ref{eqnRNADSBH}) becomes \cite{Kubiznak:2012wp}
\begin{equation}
P = \frac{T}{v} - \frac{1}{2 \pi v^2} + \frac{2q^2}{\pi v^4}
\end{equation} which resembles the equation of state of a non-ideal fluid, i.e. a fluid with interactions among its molecules. Since \(v\) is the specific volume, its reciprocal, i.e. \(\rho = 1/v\) can be interpreted as a density of the degrees of freedom \cite{Wei:2015iwa}. In terms of \(\rho\), the equation of state takes the following intuitive form \cite{meanfield2}:
\begin{equation}\label{eqnoftsaterho}
P = \rho T - \frac{\rho^2}{2 \pi} + \frac{2q^2 \rho^4}{\pi}.
\end{equation}
It may be speculated that the underlying degrees of freedom have some resemblance with those of a fluid. Noting that by the equipartition theorem, the kinetic energy of molecules is proportional to \(T\), the first term appearing in the RHS of eqn (\ref{eqnoftsaterho}) can be interpreted as a kinetic energy density of the degrees of freedom. Then, it is natural to interpret the remaining terms as a potential energy density, i.e. one defines the potential energy density:
\begin{equation}\label{u}
u (\rho) = - \bigg(\frac{1}{2 \pi}\bigg)\rho^2 + \bigg(\frac{2q^2 }{\pi}\bigg) \rho^4
\end{equation} which has been plotted in figure-(\ref{potential}).\\

If the sign of the thermodynamic curvatures indicates towards the nature of microscopic interactions, then the point of zero crossing (\(\tilde{R}_U = R_H = 0\)) is expected to coincide with the extremum of the mean field interaction potential, i.e. 
\begin{equation}
\frac{\partial u(\rho)}{\partial \rho} = 0.
\end{equation} The condition above gives \(\rho = 1/\sqrt{8}|q|\) or \(r_+ = \sqrt{2} |q|\), exactly coinciding with the point at which the thermodynamic curvatures (both \(\tilde{R}_U\) and \(R_H\)) vanish. From a fluid-like perspective, in a mean field description, where the particle positions are averaged, one can argue that \(v \sim r^3\) or \(\rho \sim r^{-3}\) where \(r\) is the mean separation between the particles. Thus, from eqn (\ref{u}), one can speculate that the microscopic potential describing the interactions among the degrees of freedom can be taken to be of the form \cite{meanfield2}:
\begin{equation}
V(r) = -\frac{\sigma}{r^6} + \frac{\delta}{r^{12}}, \hspace{5mm} \sigma, \delta > 0.
\end{equation} Therefore, the Lennard-Jones potential describes the interactions between the microscopic degrees of freedom, at least at a mean field level \cite{Wei2,meanfield2,Dutta,Weimf}.\\

The remarks made above may be generalized straightforwardly to the case of charged AdS black holes in an arbitrary number of spacetime dimensions. It should be mentioned that similar studies have been performed for black holes in higher curvature theories, such Einstein-Gauss-Bonnet theory \cite{weiGB,meanfield1,GBfixedphi,wei4DGB,mansoori4DGB}. The analysis of the thermodynamic curvatures for such systems can be done in a similar manner as described above for charged AdS black holes in Einstein gravity. Although we do not pursue it further, we summarize the sign of the thermodynamic curvature for several black holes in AdS in the extended thermodynamic framework in table-(II). In the next subsection, we shall consider an alternate set-up, where black hole chemistry has been studied in the context of the AdS/CFT correspondence. 

  \begin{table*}[t]
\caption{Sign of Ruppeiner curvature for some black holes in AdS \cite{Wei1,Wei2,Cai,Aman:2003ug,Shen:2005nu,Sarkarbtz,Mirza:2007ev,Quevedo:2007mj,Banerjee:2010da,Sahay:2010tx,Liu:2010sz,AR,Wei:2015iwa,Mansoori:2016jer,Bhattacharya:2017hfj,Miao:2017fqg,Xu,weiGB,meanfield1,wei4DGB,mansoori4DGB,btz,btz2,meanfield2,Dehyadegari:2020ebz,HosseiniMansoori:2020jrx,unicrit,NaveenaKumara:2020biu,Wei:2020kra,Yerra:2020tzg,chemical1,chemical2,chemical3,chemical4,chemical5}. In the table below, by ``neutral" we mean the absence of electric charge.\\}
\centering
\begin{tabular}{| c | c | c | c |}
\hline
{\bf Black hole}  & {\bf Sign of thermodynamic curvature}    \\ \hline  
Neutral and non-rotating BTZ         &  \(0\) \\ \hline
Rotating BTZ        &  \(+\)\\ \hline
Charged BTZ         &  \(+\)\\ \hline
Exotic BTZ         &  \(0\), $\pm$\\ \hline
Schwarzschild-AdS in d-dimensions        &  \(-\)\\ \hline
 Reissner-Nordstr\"{o}m-AdS in d-dimensions        &  \(0\), $\pm$ \\ \hline
  Neutral or charged Gauss-Bonnet-AdS in 4-dimensions       &  \(0\), $\pm$ \\ \hline
 Neutral Gauss-Bonnet-AdS in 5-dimensions       &  \(-\) \\ \hline
 Charged Gauss-Bonnet-AdS  in 5 and 6-dimensions      &  \(0\), $\pm$ \\ \hline
  Neutral Gauss-Bonnet-AdS in 6-dimensions       &  \(0\), $\pm$ \\ \hline
\end{tabular}
\end{table*}

\subsection{Boundary} \label{holosec}
A deep motivation for studying the thermodynamics of black holes in AdS, is the all too important AdS/CFT correspondence \cite{Maldacena:1997re,Witten:1998qj,Witten:1998zw}, which is a duality relating: a certain (quantum) theory of gravity in \(d\)-dimensional AdS spacetime (known as the bulk) to a conformal field theory (CFT) which is defined on the $(d-1)$-dimensional boundary. One of the remarkable checks of this correspondence is the identification of the cross-over from thermal AdS phase to the black hole phase (the Hawking-Page transition), with the large $N$ confinement-deconfinement transition in the boundary field theory \cite{Witten:1998qj}. This correspondence has of course received continuous attention with the most well studied case being the correspondence between string theory in AdS$_5 \times S^5$ and $\mathcal{N}=4$, $SU(N)$ supersymmetric Yang-Mills theory at large $N$, meaning that there exists a relation connecting parameters on both sides of the duality, namely \cite{Maldacena:1997re}:
\begin{equation}\label{adscft}
l^4 = \frac{\sqrt{2}l_{pl}^4}{\pi^2}  N \, .
\end{equation}
Here, $l$ gives a measure of number of degrees of freedom via $N$ which is the number of colors of the boundary gauge theory with $l_{pl}$ being the ten dimensional Planck length.\\

Now, with regards to extended thermodynamics motivated above, we saw the possibility of having new pressure $P$ and thermodynamic volume $V$, variables in the bulk (gravity) description. It is tempting to ask what these quantities correspond to in the holographic dual field theory via the AdS/CFT correspondence. There are several arguments which reveal that the pressure $P$ in bulk introduced as above, is not the usual pressure of the boundary field theory \cite{Johnson,DolanBose,Karch}. The pressure of the boundary theory is fully determined from the partition function of the theory. However, in the bulk, the pressure comes from a different notion of a variable $\Lambda$. It is useful to remember that the length scale $l$ comes from the underlying uncompactified theory, and is just a number, $N$, related to the number of D-branes in theory. Gauge/gravity duality is well studied in the large $N$ limit giving several clarifying results, which is the limit of large $l$ or small curvature limit. On the field theory side, $N$ is generally the rank of the gauge group and sets the number of degrees of freedom (which are actually proportional to $N^2$ for the $U(N)$ gauge group). This means that a dynamical $\Lambda$, giving pressure $P = -\Lambda/{8\pi G_d}$ in the bulk, should correspond to a dynamical $N$ on the holographic dual side \cite{DolanBose}. Varying the number of branes $N$, might mean holographic renormalization group (RG) flow and more interestingly, to a tour in the space of dual field theories \cite{Johnson}. It is well known that RG flow changes the effective cosmological constant of the underlying theory and also plays an active role in changing the number of degrees of freedom.\\

It is important to note here that: in traditional black hole thermodynamics (when $\Lambda$ is not dynamical), the gauge/gravity correspondence suggests identifying the bulk quantities such as temperature $T$ and entropy $S$ with the quantities in the boundary. What changes now, in the context of a dynamical $\Lambda$ is that, although black hole mass $M$ continues get identified with internal energy $U$ of the boundary \cite{Johnson}; in the bulk $M $ is identified with enthalpy $H= U + P V$. This holographic interpretation was argued in several works to be a plausible starting point to discuss holographic aspects of black hole heat engines, most notably in~\cite{Johnson}. There are further subtleties such as the role of the Newton's constant $G_d$ in the extended first law of black hole thermodynamics. Such questions are currently being explored \cite{Visser,Cong1,Cong2}. \\

Let us consider the approach adopted in \cite{chemical1,chemical2,chemical3,chemical4,chemical5}, where a dynamical cosmological constant in the bulk corresponds to varying the number of colours on the boundary \cite{DolanBose}. For definiteness, we shall consider black holes in AdS\(_5 \times\) S\(^5\). The bulk metric field reads \cite{Chamblin}
\begin{equation}
\label{10d}
ds^2=ds_{AdS_5}^2 + l^2\,d\Omega_{5}^2
\end{equation}
where, $d\Omega_{5}^2$ is the line element on a 5-dimensional sphere with unit radius and, 
\begin{equation} \label{bh}
 ds_{AdS_5}^2 = -f(r) \, dt^2 + f(r)^{-1} \, dr^2 + r^2\, d\Omega_3^2
\end{equation} with \(d\Omega_3^2\) being the metric on a 3-sphere. Here, \(f(r)\) is the blackening factor which has the following form:
\begin{equation}
  f(r) = 1 - \frac{8G_{(5)}M}{3\pi r^2} + \frac{r^2}{l^2}
\end{equation}
where \(l\) is the radius of the AdS\(_5\) spacetime related to the cosmological constant as $\Lambda = -\frac{6}{l^2}$. $M$ is the black hole mass and the five dimensional Newton's constant \(G_{(5)}\) appearing in the black hole solution is not fixed but is tied to $l$ as
\begin{equation} \label{bulk}
\frac{1}{16\pi G_{(5)} } = \frac{\pi^2 l^5}{16 G_{(10)}}\, .
\end{equation}
Here, it is the ten dimensional Newton's constant $G_{(10)}$ and the ten dimensional Planck length $l_P$ (linked as $\hbar G_{(10)} = l^8_P$) which are held fixed. The spacetime AdS\(_5 \times\) S\(^5\) can be thought of as the near horizon limit of \(N\) coincident D3-branes stacked on top of each other in type IIB supergravity. The dual description is the \(\mathcal{N} = 4\) \(SU(N)\) SUSY Yang-Mills theory in the large \(N\) limit. The particle content of the theory is: \(N^2\) gauge fields, \(6N^2\) massless scalars, and \(4N^2\) Weyl fermions. Thus, there are \(8N^2\) bosonic and \(8N^2\) fermionic degrees of freedom. Therefore, the number of degrees of freedom scales as \(N^2\) (proportional to the central charge). Following \cite{chemical3,chemical5}, we shall be considering energy and entropy densities instead of their absolute values for convenience. Thus, the first law reads
\begin{equation}\label{boundaryfirstlaw}
du = T ds + \mu dN^2 \, 
\end{equation} where \(u = 2 \pi^2 M/\mathcal{V}\) is the energy density, \(s = 2\pi^2 S/\mathcal{V}\) is the entropy density, and \(\mu\) is the chemical potential for the number of degrees of freedom. Here \(\mathcal{V} = 2 \pi^2 l^3\) is the CFT volume and the AdS radius \(l\) is related to the number of D3-branes via eqn (\ref{adscft}). In particular, the energy density in terms of thermodynamic quantities is given by
\begin{equation} \label{rhodensity}
  u = \frac{3 s^{2/3} \left(N^{5/6}+2 \sqrt[3]{2} s^{2/3}\right)}{4\ 2^{2/3} \pi  N^{2/3}}.
\end{equation} Thus, from eqn (\ref{boundaryfirstlaw}), the Hawking temperature is calculated to be
\begin{equation} \label{Tdensity}
T= \bigg(\frac{\partial u}{\partial s}\bigg)_{N^2} =\frac{N^{5/6}+4 \sqrt[3]{2} s^{2/3}}{2 \times 2^{2/3} \pi  N^{2/3} \sqrt[3]{s}}
\end{equation}
which has a minimum value, \(T_{\rm min}\) at \(s_0 =\frac{N^{5/4}}{8 \sqrt{2}}\). For any temperature above \(T_{\rm min}\), there are two values of \(s\) with the same temperature: \(s < s_0\) corresponds to the small black hole branch whereas \(s > s_0\) corresponds to the large black hole branch \cite{Hawking:1982dh}. From the energy density [eqn (\ref{rhodensity})] one can compute the chemical potential for \(N^2\), which reads
\begin{equation} \label{mudensity}
  \mu =: \bigg(\frac{\partial u}{\partial N^2}\bigg)_s = \frac{\sqrt[3]{2} N^{5/6} s^{2/3}-8 \times 2^{2/3} s^{4/3}}{32 \pi  N^{8/3}}\, .
\end{equation}
Before computing the thermodynamic curvatures, it is important to find the specific heats. \(C_{N^2}\) is given by
\begin{equation}\label{CN2}
  C_{N^2} = \bigg(\frac{\partial u}{\partial T}\bigg)_{N^2} = -\frac{3 s \left(N^{5/6}+4 \sqrt[3]{2} s^{2/3}\right)}{N^{5/6}-4 \sqrt[3]{2}
   s^{2/3}} \, .
\end{equation}
It diverges at $T = T_{\rm min}$ and is positive in the region of the large black hole while it is negative for that of the small black hole~\cite{Hawking:1982dh}. In a fixed chemical potential setting, \(C_\mu\) is calculated to be
\begin{equation}\label{cmu}
C_{\mu} = \bigg(\frac{\partial h}{\partial T}\bigg)_{\mu} = -\frac{-\frac{512\ 2^{2/3} s^{7/3}}{N^{5/6}}+11 N^{5/6} s-84 \sqrt[3]{2}
   s^{5/3}}{3 N^{5/6}-36 \sqrt[3]{2} s^{2/3}}
\end{equation} where \(h = u - \mu N^2\). Thus, we now have two ensembles, one with fixed \(N^2\) while the other is with fixed \(\mu\). In the latter, the first law reads: \(dh = Tds - N^2 d\mu\). \\

\begin{figure}[t]
\begin{center}
\includegraphics[width=3.2in]{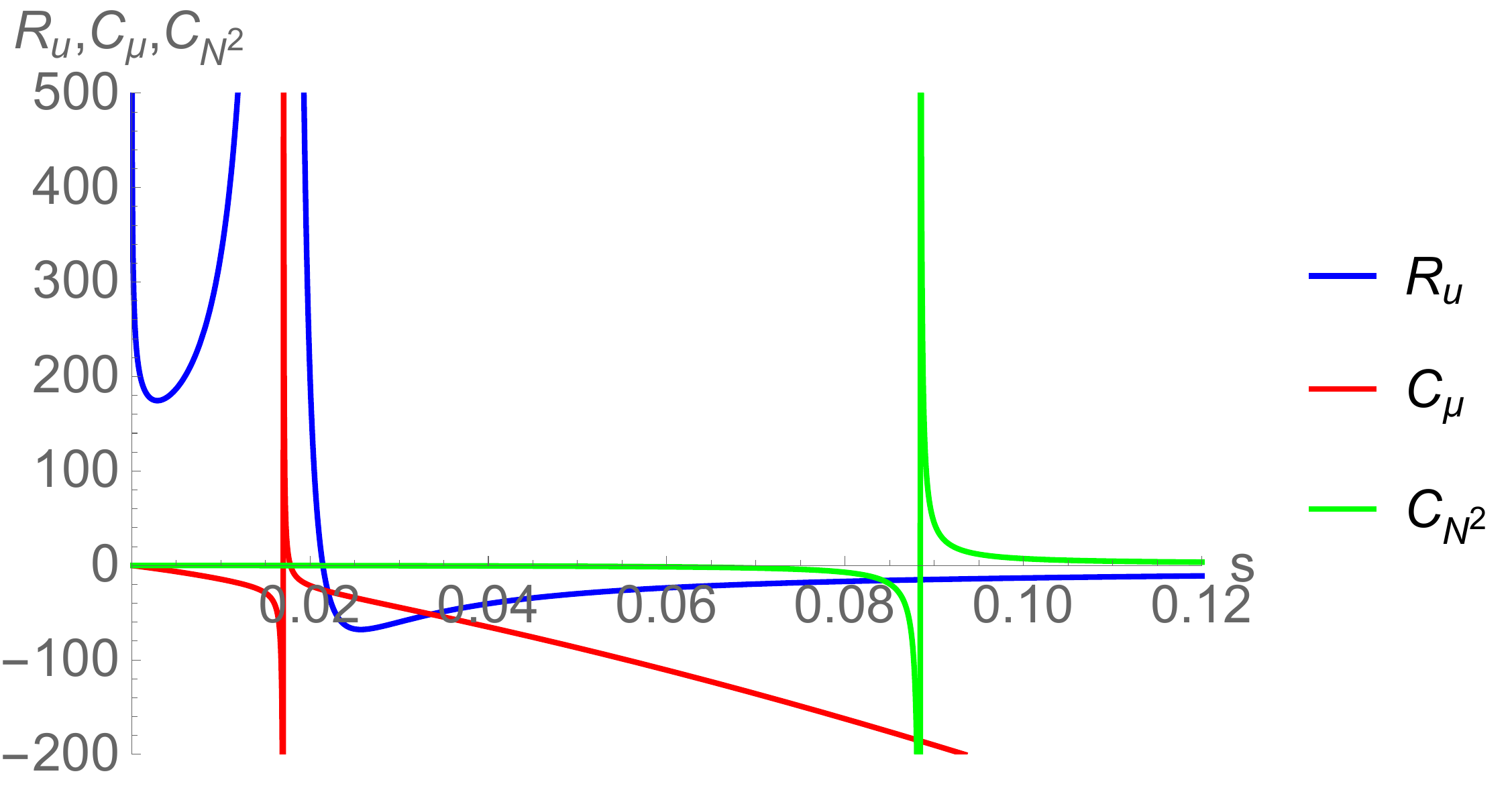}
\caption{Plot of thermodynamic curvature \(R_u\) together with the specific heats \(C_{N^2}\) and \(C_\mu\) as a function of entropy density \(s\).}
\label{RuCmuCN}
\end{center}
\end{figure}

Let us begin with the fixed \(N^2\) ensemble, in which following the treatment presented in section-(\ref{section3}), we have
\begin{equation}
dl^2_R = \frac{1}{T} (ds \otimes dT + dN^2 \otimes d\mu)
\end{equation} and this can be written in different parametrizations. The associated thermodynamic curvature, \(R_u\) takes the following form:
\begin{equation}\label{Rsn}
R_u = \frac{C_1}{C_2},
\end{equation} where,
\begin{eqnarray}
  C_1= 8 (40\ 2^{2/3} N^{5/3} s^{2/3}+160 N^{5/6} s^{4/3} \nonumber \\
     -5 \sqrt[3]{2} N^{5/2}+768 \sqrt[3]{2} s^2), \nonumber \\
  C_2 = 3 N^{5/6} \sqrt[3]{s} \left(N^{5/6}-12 \sqrt[3]{2} s^{2/3}\right)^2 \nonumber \\
   \times \left(N^{5/6}+4 \sqrt[3]{2} s^{2/3}\right). \nonumber
\end{eqnarray}
The thermodynamic curvature \(R_u\) has been plotted in figure-(\ref{RuCmuCN}), together with the specific heats \(C_\mu\) and \(C_{N^2}\). Clearly, the divergence of \(R_u\) coincides with that of \(C_\mu\) as one would have expected. Furthermore, if \(T_0\) be the temperature at which \(R_u\) and \(C_\mu\) diverge (as shown in figure-(\ref{RuCmuCN})), it is straightforward to show \cite{chemical5}:
\begin{equation}\label{cmuexpon}
C_\mu \sim |T - T_0|^{-1}, \hspace{4mm} R_u \sim |T - T_0|^{-2}
\end{equation} thereby giving the same exponents as found in the bulk concerning the divergence of \(C_P\) and the normalized curvature \(\tilde{R}_U\) respectively. The same exponents have been observed in various different contexts earlier \cite{exponent,information,Wei1,unicrit,chemical5} including the case of black holes in the bulk. \\
  \begin{figure}[t]
\begin{center}
\includegraphics[width=3.2in]{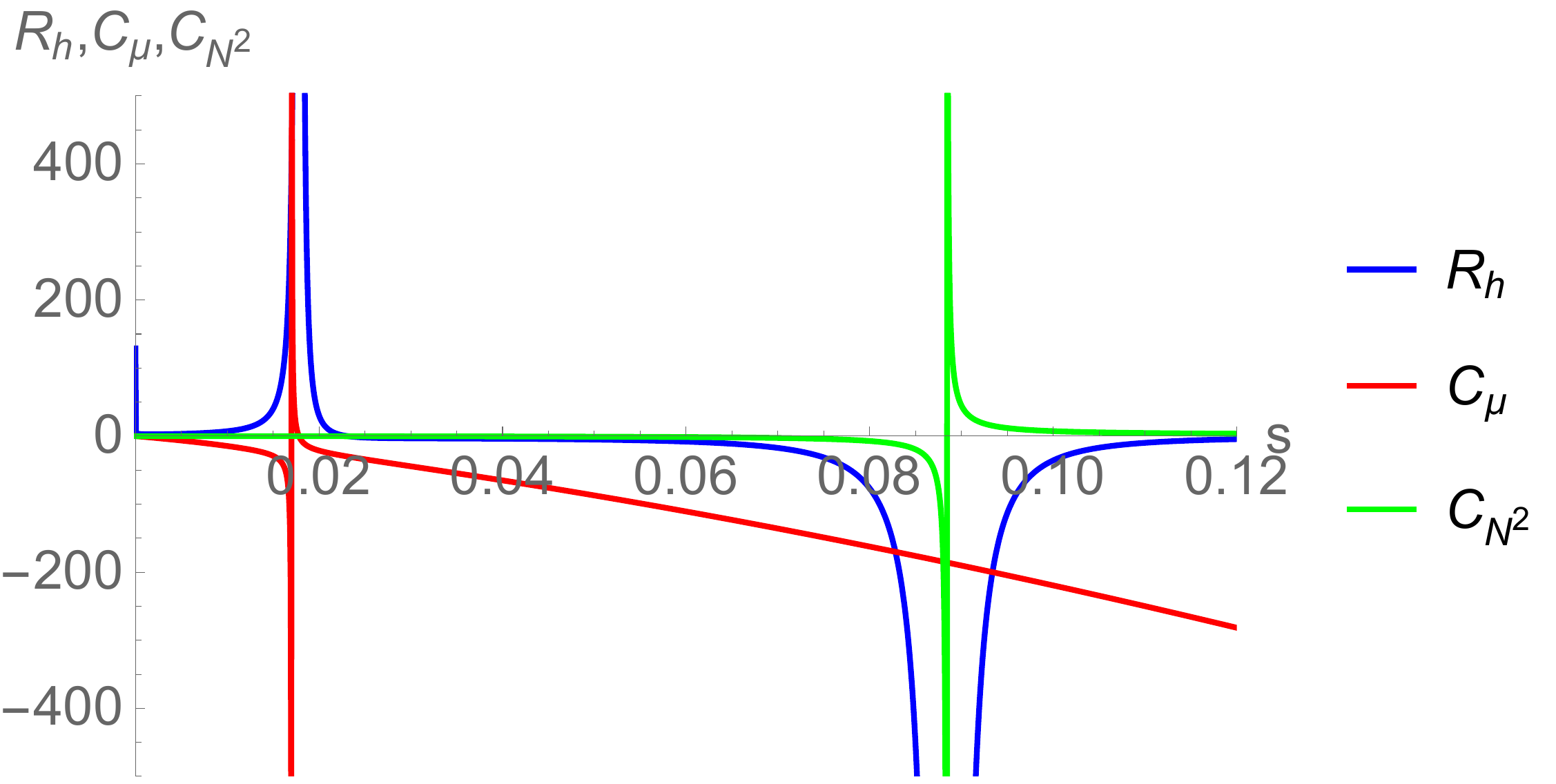}
\caption{Plot of thermodynamic curvature \(R_h\) together with the specific heats \(C_{N^2}\) and \(C_\mu\) as a function of entropy density \(s\).}
\label{RhCmuCN}
\end{center}
\end{figure}

In the fixed \(\mu\) ensemble, the Ruppeiner metric turns out to have the following form:
\begin{equation}
dl^2_R = \frac{1}{T} (ds \otimes dT - dN^2 \otimes d\mu).
\end{equation} The associated curvature scalar, labelled as \(R_h\) is computed to be
\begin{equation}
R_h=\frac{D_1}{D_2 D_3}
\end{equation} 
where,
\begin{eqnarray}
D_1 &= &-12 N^{5/6} \left(-968\ 2^{2/3} N^{5/3} s^{2/3}-6368 N^{5/6}
   s^{4/3} \right.  \nonumber \\ 
     &+ &  \left. 165 \sqrt[3]{2} N^{5/2}-61440 \sqrt[3]{2} s^2\right), \nonumber \\
  D_2 &=& \sqrt[3]{s} \left(N^{5/6}+4 \sqrt[3]{2} s^{2/3}\right),  \nonumber \\
  D_3 &=&\left(-172 \sqrt[3]{2} N^{5/6} s^{2/3}+11 N^{5/3}+512\ 2^{2/3}
   s^{4/3}\right)^2 . \nonumber
  \end{eqnarray}
  The curvature scalar \(R_h\) has been plotted in figure-(\ref{RhCmuCN}), together with the specific heats  \(C_{N^2}\) and \(C_\mu\). In contrast to \(R_u\), the thermodynamic curvature \(R_h\) does not diverge at the divergence of \(C_\mu\). However, its divergence coincides with that of \(C_{N^2}\) with the same exponents as eqn (\ref{cmuexpon}) (with \(C_\mu\) replaced by \(C_{N^2}\), and \(R_u\) replaced by \(R_h\)).  Furthermore, one may observe that \(R_h\) diverges at the zero of \(C_\mu\). This is precisely what we expect based on the discussions presented in section-(\ref{section3}). Henceforth, we have demonstrated the generality of the arguments presented in section-(\ref{section3}) and the correspondence between the divergences of specific heats and thermodynamic curvatures in different ensembles related by Legendre transforms [table-(I)].  \\
  
  Now, if one considers the sign of the thermodynamic curvature to be an empirical indicator of the nature of microscopic interactions, then clearly for the large black hole branch ($s > 0.0883883$) one has \(R_u, R_h < 0\) suggesting that the system is attraction dominated, reminiscent of an ideal gas of bosons. Moreover, it was shown in \cite{chemical5} (see also \cite{DolanBose}) that \(|R_u|\) increases as one approaches towards \(z \rightarrow 1\) where \(z = e^{\mu/T}\) is the fugacity parameter. For an ideal gas of bosons, this limit indicates Bose condensation wherein the absolute value of the thermodynamic curvature grows indicating the growth of inter-particle correlations \cite{quantum}. The fact that the same behavior is observed for black holes in AdS\(_5 \times S^5\) may suggest that the degrees of freedom undergo an analogous condensation \cite{DolanBose}. However, a satisfactory understanding of this can only be achieved via computations performed in a quantum theory of gravity. Nevertheless, the study of the thermodynamic curvature may reveal early insights into the physics of black holes.

\section{Discussion}\label{section5}
Geometrical approaches to thermodynamics and in particular, thermodynamics of black holes have received constant attention due to their potential to provide a unique perspective on connecting the microscopic to macroscopic physics \cite{Wei1,Wei2,Cai,Aman:2003ug,Shen:2005nu,Sarkarbtz,Mirza:2007ev,Quevedo:2007mj,Hendi:2015rja,Mansoori:2016jer,Banerjee:2010da,Sahay:2010tx,Liu:2010sz,AR,Wei:2015iwa,Bhattacharya:2017hfj,Miao:2017fqg,Xu,weiGB,meanfield1,wei4DGB,mansoori4DGB,btz,btz2,meanfield2,Dehyadegari:2020ebz,HosseiniMansoori:2020jrx,unicrit,NaveenaKumara:2020biu,Wei:2020kra,Yerra:2020tzg,chemical1,chemical2,chemical3,chemical4,chemical5}. As summarised in this review, methods of contact and metric geometry have given novel insights (though qualitative in nature) on the nature of dominant interactions and phase transitions in black holes in AdS in the extended thermodynamics set up. It should be mentioned here that the thermodynamic metrics explored in this review have been generalized further by several groups with varied advantages, such as ~\cite{geometrotherm,Quevedo:2007mj,Hendi:2015rja,Mansoori:2016jer}, among others. For instance, in the framework of geometrothermodynamics \cite{geometrotherm,Quevedo:2007mj}, the thermodynamic metric is Legendre invariant, i.e. it is invariant under Legendre transformations. However, the metric is not a Hessian although there have been recent attempts to derive it from statistical mechanics \cite{stator}. \\

In this review, we considered Hessian thermodynamic metrics in different ensembles connected by (partial) Legendre transforms and discussed their complimentary behavior as far as divergences are concerned \cite{bravettiensemble}. While such metrics are not Legendre invariant, they are physically straightforward to motivate on the grounds of thermodynamic fluctuation theory. We have emphasized upon ensemble non-equivalence and reparametrizations of Hessian metrics in various choices of independent coordinates. We then considered the most widely used Hessian metric, the Ruppeiner metric \cite{Rup1,Rup2} and listed the sources of its divergences from general considerations. They were then verified through various examples considered subsequently. It was mentioned that the sign of the thermodynamic curvature could possibly indicate towards the nature of microscopic interactions in a thermodynamic system. While this can indeed be verified for the van der Waals fluid or ideal quantum gases \cite{quantum}, one cannot yet ascertain its validity for a general thermodynamic system. However, keeping in mind that black holes in the extended thermodynamics framework do admit a van der Waals-like behavior, one may gain early insights into the microscopic interactions from studying the behavior of the thermodynamic curvature. In this sense, it is encouraging to explore the thermodynamic geometry of black holes in various settings. \\

In section-(\ref{bh1}), we applied the ideas developed in sections-(\ref{section2}) to (\ref{section3}), to study thermodynamic geometry of black holes in AdS spacetimes in the extended thermodynamics framework. In subsections-(\ref{bulk1}) and (\ref{holosec}), the thermodynamic geometries of the bulk and the boundary (via the gauge/gravity duality) settings were discussed respectively. We briefly touched upon the applications of thermodynamic geometry in a holographic setting where the black hole in the AdS bulk is dual to a finite temperature gauge theory on the boundary. While some consistent results were demonstrated including the exponent `2' for the thermodynamic curvature, it should be pointed out that in the context of extended thermodynamics, there have been recent developments on new ideas in relating the holographic dual theories \cite{Visser,Cong1,Cong2}. It would be interesting to extend the methods summarized in this review to such situations.

\section*{Acknowledgements}
A.G. would like to thank the Ministry of Education (MoE), Government of India for financial support in the form of a Prime Minister's Research Fellowship (ID: 1200454). C.B. gratefully acknowledges the support received from DST (S.E.R.B.), Government of India, MATRICS (Mathematical Research Impact Centric Support) grant no. MTR/2020/000135.

\end{document}